\documentclass[aps,prc,preprint,showpacs,superscriptaddress,amsmath,amssymb,nofootinbib,floatfix]{revtex4}

\usepackage[dvips]{graphicx,color}
\usepackage{dcolumn}
\usepackage{bm}

\def\pmb#1{\setbox0=\hbox{#1}%
  \kern-.025em\copy0\kern-\wd0 
  \kern.05em\copy0\kern-\wd0
  \kern-.025em\raise.0433em\box0 }

\newcommand{\Figuretable}[1]{%
  \begin{center} --------- {\bf #1} --------- \\ \end{center}} 
\def\lambdabar{\protect\@lambdabar}
\def\@lambdabar{%
\relax
\bgroup
\def\@tempa{\hbox{\raise.73\ht0
\hbox to0pt{\kern.25\wd0\vrule width.5\wd0
height.1pt depth.1pt\hss}\box0}}%
\mathchoice{\setbox0\hbox{$\displaystyle\lambda$}\@tempa}%
{\setbox0\hbox{$\textstyle\lambda$}\@tempa}%
{\setbox0\hbox{$\scriptstyle\lambda$}\@tempa}%
{\setbox0\hbox{$\scriptscriptstyle\lambda$}\@tempa}%
\egroup
}

\begin{document}

\preprint{}

\title{Deeply-bound $\bm{K^- pp}$ state 
in the $^3$He(in-flight $\bm{K^-}$,~$\bm{n}$) spectrum \\
and its moving pole near the $\bm{\pi \Sigma N}$ threshold
}
\author{Takahisa Koike\footnote
{E-mail:tkoike@riken.jp}
}%
\affiliation{%
Advanced Meson Science Laboratory, 
RIKEN Nishina Center, Wako-shi, Saitama 351-0198, Japan
}

\author{Toru Harada\footnote
{E-mail:harada@isc.osakac.ac.jp}
}%
\affiliation{%
Research Center for Physics and Mathematics,
Osaka Electro-Communication University, Neyagawa, Osaka, 572-8530, Japan
}

\date{\today}

\begin{abstract}
The formation of a deeply-bound $K^- pp$  state with $I=1/2$, $J^\pi=0^-$ by 
the $^3$He(in-flight $K^-$, $n$) reaction is theoretically investigated 
in a distorted-wave impulse approximation using the Green's function method. 
The expected inclusive and semi-exclusive spectra at 
$p_{K^-} = 1.0$ GeV/c and $\theta_{\rm lab} = 0^{\circ}$ are calculated 
for the forthcoming J-PARC E15 experiment. 
We demonstrate these spectra with several phenomenological
$K^-$-``$pp$'' optical potentials $U^{\rm opt}(E)$ 
which have an energy-dependent imaginary part multiplied 
by a phase space suppression factor, 
fitting to recent theoretical predictions or 
experimental candidates of  
the $K^-pp$ bound state. 
The results show that a cusp-like peak at the $\pi \Sigma N$ threshold 
is an unique signal for the $K^-pp$ bound state
in the spectrum including the [$K^-pp$] $\to$ $Y + N$ decay process 
from the two-nucleon $K^-$ absorption, 
as well as a distinct peak of the $K^-pp$ bound state.
The shape of the spectrum is explained by 
a trajectory of a moving pole of the $K^-pp$ bound state 
in the complex energy plane. 
The importance of the [$K^-pp$] $\to$ $Y + N$ spectrum
is emphasized in order to extract 
clear evidence of the $K^-pp$ bound state.

\end{abstract}


\maketitle

\section{Introduction}

The antikaon-nucleon ($\bar{K}N$) interaction in nuclei
is very important to elucidate the nature of high dense 
nuclear matter \cite{KBL96}. Since the $\bar{K}N$ $I=$ 0 
interaction is believed to be strongly attractive, 
one would expect the existence of deeply-bound $\bar{K}$ 
nuclear states~\cite{Dote04}.
Especially, a three-body ${\bar K}NN$ (unstable) bound state 
with a $[{\bar K}\otimes \{NN \}_{I=1}]_{I=1/2}$, $J^\pi=0^-$ 
configuration, which is called ``$K^-pp$'' symbolically, 
is suggested to be the lightest and the most fundamental 
${\bar K}$ nucleus.

In 1963, Nogami~\cite{Nog63} firstly discussed a possible 
existence of the $K^-pp$ bound state by a rather crude calculation. 
About 40 years later, Yamazaki and Akaishi~\cite{YA02} restarted 
to study the structure of the $K^-pp$ bound state based on 
a quantitative few-body calculation with a phenomenological 
$\bar{K}N$ interaction which reproduces the mass and width of 
$\Lambda(1405)$ as a $\bar{K}\text{--}N$ quasibound state. 
They predicted that the binding energy and 
width for the $K^-pp$ state are $B.E. = 48$ MeV 
and ${\varGamma} = 61$ MeV, respectively. 
Many other theoretical 
works~\cite{She07,Ike07,Dot08,Iva05,Nis08,Ara08,Yam08,Wyc08} 
have also supported the existence of the $K^-pp$ bound state, 
but the predicted binding energies and widths are not converged
(see Fig.~\ref{fig:2}); 
Shevchenko, Gal and Mare$\breve{\rm s}$~\cite{She07} performed 
a $\bar{K}NN\text{--}\pi \Sigma N$ coupled-channel Faddeev calculation
using phenomenological $\bar{K}N\text{--}\pi \Sigma$ interactions, 
leading to $B.E.$ = $55\text{--}70$ MeV and 
${\varGamma}$ = $95\text{--}110$ MeV.
Ikeda and Sato~\cite{Ike07} also obtained $B.E. = 79$ MeV and 
${\varGamma} = 74$ MeV in a similar Faddeev calculation 
with a Chiral SU(3) based $\bar{K}N\text{--}\pi Y$ interaction. 
On the other hand, some authors claimed
$B.E. \simeq 20$ MeV~\cite{Dot08,Yam08} with $\bar{K}N$ 
interactions based on the Chiral unitary approach~\cite{Ose98,Hyo08}, 
which are less attractive than the phenomenological $\bar{K}N$ ones
in the bound state region.
The discrepancy between theoretical results perhaps come from 
an ambiguity of the $\bar{K} N$ interaction,
together with a different procedure for a three-body calculation 
involving decay processes.
Further theoretical investigations are 
apparently needed.

Recently, several experimental observations of the $K^-pp$ 
state have been reported: 
The data by the FINUDA collaboration at DA$\Phi$NE~\cite{FINUDA05}
suggested evidence of a deeply-bound $K^-pp$ 
state in invariant-mass spectroscopy from stopped $K^-$ 
reactions on $^6$Li, $^7$Li and  $^{12}$C. 
Their measured energy and width are $B.E.=$ 115$\pm$9 MeV 
and ${\varGamma}$ = 67$^{+16}_{-14}$ MeV, respectively, 
whereas Magas $et$ $al.$~\cite{Mag06} claimed 
that the FINUDA data can be explained 
without postulating the existence of the $K^-pp$ bound state. 
The OBELIX experiment at LEAR-CERN~\cite{OBELIX07} also 
suggested the observation of the $K^-pp$ state
in invariant-mass spectroscopy from stopped $\bar{p}$ 
reactions on $^4$He. 
Very recently, Yamazaki $et$ $al.$~\cite{DISTO08} have found  
new experimental events of the $K^-pp$ state 
in $p + p \to K^+ + \Lambda + p$ reactions
by a reanalysis of old DISTO experimental data
at SATURNE-Saclay. 
However, these experimental results would also leave room
for other interpretations and therefore
more experimental data are required in order to confirm 
whether the $K^-pp$ system has a deeply-bound state or not.

Iwasaki $et$ $al.$~\cite{Iwa06} have proposed 
a new experiment searching the deeply-bound $K^-pp$ state at J-PARC 
by the missing-mass spectrum of the $^3$He(in-flight $K^-$, $n$) 
reaction, together with invariant-mass spectra detecting all particles 
via decay processes from the $K^-pp$ bound state 
(J-PARC E15 experiment).
Moreover, a measurement of the $K^-pp$ state in $p + p$ collisions 
has been planned by the FOPI collaboration at GSI~\cite{FOPI08}, 
as proposed by Yamazaki and Akaishi~\cite{Aka07}.
A search for the light $\bar{K}$ nuclear systems involving the 
$K^-pp$ state in the stopped $K^-$ reactions on $^3$He/$^4$He targets
has been also planned by the AMADEUS collaboration 
at DA$\Phi$NE~\cite{AMAD08}.

Our purpose is to theoretically clarify the expected inclusive 
and semi-exclusive spectra by the $^3$He(in-flight $K^-$, $n$) 
reaction for the forthcoming J-PARC E15 experiment.
In a previous work~\cite{Koi07}, 
we examined these spectra of the $^3$He(in-flight $K^-$, $n$) 
reactions within a distorted-wave impulse approximation (DWIA) 
employing the Green's function method~\cite{Mor94} which well describes 
unstable hadron systems~\cite{Har98}.
It has been shown that the $^3$He(in-flight $K^-$, $n$) reaction 
provides a promising spectrum which contains a 
$s$-wave dominance in the $K^-$ bound region, 
where a strong nuclear distortion for $K^-$ is reduced~\cite{Koi07}. 
This is a major advantage of a use of the $s$-shell 
nuclear target such as $^3$He.
A similar calculation for the $^3$He target is also presented 
by Yamagata $et$ $al.$~\cite{Yam08} with the help of the Chiral 
unitary approach.
The investigations for heavier targets 
within a similar framework have been found in several 
publications~\cite{Kis05,Iku02,Yam06,Koi08}.
In the case of $p$-shell targets of $^{12}$C and 
$^{16}$O~\cite{Kis05}, 
the signals of the $\bar{K}$ nuclear states would not be extracted 
due to their broad widths even
if the bound states exist~\cite{Yam06,Koi08}.

In this paper, we theoretically investigate the formation and decay 
of the deeply-bound $K^- pp$ state by 
the $^3$He(in-flight $K^-$,~$n$) reaction 
at the incident $K^-$ momentum $p_{K^-}=$ 1.0 GeV/c and 
the forward direction $\theta_{\rm lab}=$ 0$^\circ$ 
within a DWIA. 
To search a signal of the deeply-bound $K^-pp$ state, 
we examine the inclusive and semi-exclusive spectra 
including one-nucleon $K^-$ absorption processes,
%
\begin{eqnarray}   
[\, K^-pp \, ]  \to  ``K^-p"+p & \to & \pi + Y + N, 
\label{eq:decaymode1}
\end{eqnarray}
and two-nucleon $K^-$ absorption processes, 
%
\begin{eqnarray}   
[\, K^-pp \, ] & \to  K^- + ``pp"   & \to Y + N, 
\label{eq:decaymode2}
\end{eqnarray}
near the $\pi \Sigma N$ decay threshold, 
where $Y=\{\Sigma, \Lambda\}$.
Since there exist many predictions of $B.E.$ and $\varGamma$ for 
the $K^-pp$ state at present, 
we demonstrate typical spectra by using the $K^-$-``$pp$" optical 
potential which reproduces the values of each $B.E.$ 
and $\varGamma$ phenomenologically. 
Here we employ the phenomenological $K^-$-``$pp$" optical potentials 
having an energy-dependence due to the phase space factors of 
the above processes (\ref{eq:decaymode1}) and (\ref{eq:decaymode2}).
If $B.E.$ is larger than about $100$ MeV,
a decay channel via the process (\ref{eq:decaymode1}) 
with $Y = \Sigma$ is kinematically closed, so that
the decay width of the $K^-pp$ bound state would be 
considerably small.
Indeed, recent Faddeev calculations~\cite{She07,Ike07} 
or several experimental observations~\cite{FINUDA05,DISTO08}
have suggested that $B.E.$ is close to the energy at the $\pi \Sigma N$ decay 
threshold. In order to deal with such a threshold effect, 
we must take into account the energy dependence in the 
$K^-$-``$pp$'' optical potential. 
This is a natural extension of the previous 
work~\cite{Koi07} where we mainly discussed the spectra with 
an energy-independent optical potential.
A preliminary result of this subject is partially found in 
Ref.~\cite{Koi08-2}.

The outline of this paper is as follows:
In Sect.~\ref{DWIA}, we mention our DWIA framework 
using the Green's function method for 
the $^3$He(in-flight $K^-$,~$n$) reaction, and 
we introduce several phenomenological energy-dependent 
$K^-$-``$pp$'' optical potentials, of which parameters are
determined to reproduce the values of $B.E.$ 
and $\varGamma$ obtained from a recent few-body 
calculation~\cite{YA02,She07,Dot08} or experimental 
candidate~\cite{FINUDA05,OBELIX07,DISTO08}.
In Sect.~\ref{Results}, 
we show the calculated inclusive and semi-exclusive spectra 
with each optical potential. 
We find a distinct peak structure or a cusp-like structure 
at the $\pi \Sigma N$ threshold in the spectrum 
depending on the potential parameters; 
its shape behavior of the spectrum is governed 
by a pole trajectory 
for the $K^-pp$ state in the complex energy plane. 
In Sect.~\ref{Discuss}, we discuss the dependence 
of the spectral shape on the potential parameters 
and the branching ratio of $K^-$ absorptions 
systematically,
in order to understand the appearance of a cusp-like spectrum.
The summary and conclusion are given in Sect.~\ref{Summary}.

\section{\label{DWIA}
Framework}

\subsection{Distorted-wave impulse approximation (DWIA)}

In the DWIA framework~\cite{Huf74},
the inclusive double-differential cross 
section of the $^3$He(in-flight $K^-$,~$n$) reaction 
at the forward direction $\theta_{\rm lab}=$ 0$^\circ$
in the lab system is written~\cite{Tad95} as
\begin{eqnarray}   
\frac{d^2 \sigma}{d E_{n} d \Omega_{n}} &=&
\beta(0^\circ) \,
\left\langle \frac{d \sigma}{d \Omega_{n}}(0^\circ) \right\rangle
_{\rm lab}^{K^- N \to N  \bar{K}} \,
S(E),
\label{eq:DWIA}
\end{eqnarray}
where 
$S(E)$ is a strength function for the $K^-pp$ system
as a function of the energy $E$, 
and $\langle d \sigma / d \Omega_{n} (0^\circ) \rangle
_{\rm lab}^{K^- N \to N  \bar{K}}$ is
a Fermi-averaged cross section for the elementary 
$K^- + N \to N + \bar{K}$ forward scattering which is equivalent 
to a backward $\bar{K}+N$ elastic scattering
in the lab system~\cite{Kis99}. 
The lab cross section for the non-spin-flip 
$K^- + n \to n + K^-$ ($K^- + p \to n + \bar{K}^0$) process
amounts to 24.5 mb/sr (13.1 mb/sr) in free space~\cite{Kis99,Gop77}, 
and is reduced to 13.9 mb/sr (7.5 mb/sr) 
with Fermi-averaging~\cite{Ros80,Koi08}. 
Both of the $K^- + n \to n + K^-$ elastic scattering and 
the $K^- + p \to n + \bar{K}^0$  charge exchange reaction
can contribute to the formation of 
the $K^-pp$ $I=$ 1/2 state through
the coupling between $K^-pp$ and $\bar{K}^0pn$ channels. 
Thus an incoherent sum of 
contributions from these  $K^- + n \to n + K^-$ and 
$K^- + p \to n + \bar{K}^0$  processes,
as is done in Ref.~\cite{Yam08}, may be unsuitable for 
the $K^-pp$ bound region.
In the $\bar{K}^0$/$K^-$ charge basis,
the coupled-channel calculation would be needed.

Here we consider the cross section of Eq.~(\ref{eq:DWIA}) 
in the isospin basis because 
total isospin $I= 1/2$ is expected to be an almost 
good quantum number
in the $K^-pp$ bound state. 
The contribution from the elementary
processes is approximately estimated 
by the isoscalar $\Delta I=$ 0 transition amplitude
$f_{\Delta I=0} = -\sqrt{2/3} (f_{K^-n \to nK^-} + 
{1 \over 2}f_{K^- p \to n {\bar K}^0})$ 
including a spectroscopic factor for the 
$K^-pp$ $I$=1/2 state formed on $^3$He~\cite{Koi08}.
If we use the amplitude $f_{\Delta I=0}$ with Fermi-averaging 
at $p_{K^-}$=1.0 GeV/c and $\theta_{\rm lab}=$ 0$^\circ$, 
the Fermi-averaged cross section 
$\langle d \sigma / d \Omega_{N} (0^\circ) \rangle
_{\rm lab}^{\Delta I=0}=|\langle f_{\Delta I=0} \rangle|^2$
is found to be 16.4 mb/sr, 
whereas 1.2 mb/sr for the isovector $\Delta I=$ 1 
transition~\cite{Koi08}. 
In our calculations, therefore, we adopt 16.4 mb/sr as a value of  
$\left\langle d \sigma/d \Omega_{n}(0^\circ)
 \right\rangle_{\rm lab}^{K^-N \to N{\bar K}}$ 
in Eq.~(\ref{eq:DWIA}).

The kinematical factor $\beta(0^\circ)$~\cite{Tad95,Koi08} in Eq.(\ref{eq:DWIA})
expresses the translation from the two-body $K^-$-$n$ lab system 
to the $K^-$-$^3$He lab system at $\theta_{\rm lab}=0^\circ$~\cite{Dov83}, 
and it is defined as 
%
\begin{eqnarray}   
\beta(0^\circ) &=& \left(  1 - \frac{E_n^{(0)}}{E_{K^-}^{(0)}}
\frac{p_{K^-}-p_n^{(0)}}{p^{(0)}_n} \right)
\frac{p_n \, E_n}{p_n^{(0)} E_n^{(0)}},
\label{eq:beta}
\end{eqnarray}
where $p_{K^-}$ and $p_n$ ($E_{K^-}$ and $E_n$) are
momenta of the incident $K^-$ and the emitting $n$ (energies
of the residual ${K^-}$ and the emitting $n$ in the final state) 
in the many-body 
$K^-$ + ${^3{\rm He}}$ $\to$ $n$ + [$K^-pp$] reaction, 
respectively, and the quantities with an (0) superscript are 
in the two-body $K^- + n \to n + K^-$ reaction.
Note that the momentum transfer of this reaction becomes
negative; $q(0^\circ) \equiv p_{K^-} - p_n < 0 $.
For the negative momentum transfer,
$\beta(0^\circ)$ enhances the spectrum of Eq.(\ref{eq:DWIA})
by a factor $1\text{--}2$ depending on $p_n$ and 
$E_{K^-}$~\cite{Koi08}.

The strength function of the $K^-pp$ system, $S(E)$, 
in Eq.~(\ref{eq:DWIA}) can be given as 
a function of the energy $E$ measured from the $K^- + p +p$ 
threshold;
\begin{eqnarray}   
E &=& M_{K^-pp}-(M_{K^-}+M_{p}+M_{p}),
\label{eq:energy}
\end{eqnarray}
where $M_{K^-pp}$, $M_{K^-}$ and $M_{p}$ are the masses of 
the $K^-pp$ bound state, the $K^-$ and the proton, respectively.
In this calculations, we assumed a ``$pp$'' pair 
as a rigid core with a $^1S_0$ state. 
This assumption would be suitable for qualitatively describing 
the structure of the $K^-pp$ state, 
as long as we consider the deeply-bound region. 
A simple (1s)$^3$ harmonic oscillator model is used 
for the $^3$He wave function, 
in which the relative $2N$-$N$ wave function 
has the form of $\phi_{2N\mbox{-}N}(r) \propto \exp(-r^2/2a^2)$
where $a = b_{N} \sqrt{3/2}$. The size parameter $b_{N}$ is taken to be 
1.30 fm, which reproduces the experimental r.m.s charge radius of 
$^3$He, $\sqrt{\langle r^2 \rangle} = 1.94$ fm~\cite{Ang04}.

\subsection{Green's function method}

Here we consider Green's function for the $K^-pp$ system.
It is obtained 
by solving the Klein-Gordon equation numerically; 
%
\begin{eqnarray}   
\{ ( E - V_{\rm Coul}({\bm r}) )^2  
 + {\bm \nabla}^2 - \mu^2 - 2 \, \mu \, 
   U^{\rm opt}(E; {\bm r}) \}
\, G(E;{\bm r},{\bm r}') 
&=& \, \delta^3({\bm r}-{\bm r}'),
\label{eq:KG-Green}
\end{eqnarray}
where $\mu$ is the reduced mass between the $K^-$ and the ``$pp$'' 
core-nucleus, and $V_{\rm Coul}$ is the Coulomb potential 
with the finite nuclear size effect. 
$U^{\rm opt}(E)$ is an energy-dependent 
$K^-$-``$pp$'' optical potential between the $K^-$ and 
the ``$pp$'' core-nucleus, which is assumed to be 
the Lorentz scalar type.

According to the Green's function method~\cite{Mor94}, we 
can write $S(E)$ as
%
\begin{eqnarray}   
S(E) &=& 
- \, \frac{1}{\pi} \  {\rm Im} \left[ \  \sum_{\alpha, \alpha'} 
  \int d{\bm r} d{\bm r'} 
  f^{\dagger}_{\alpha}({\bm r}) 
  G_{\alpha, \alpha'}(E;{\bm r},{\bm r'})
  f_{\alpha'}({\bm r'}) \ \right]
\label{eq:S(E)}
\end{eqnarray}
with 
%
\begin{eqnarray}   
f_{\alpha}({\bm r}) &=& 
\chi^{(-)*} \left( {\bm p}_{n},
 \frac{M_{\rm C}}{M_{K^-pp}}{\bm r} \right)    
\ 
\chi^{(+)} \left( {\bm p}_{K^-},
 \frac{M_{\rm C}}{M_{\rm A}}{\bm r} \right)
\ 
\langle \alpha \, | \psi_n ({\bm r}) | \,  i \rangle,
\label{eq:chi+-}
\end{eqnarray}
where $G_{\alpha, \alpha'}(E)$ is the complete Green's function for the 
$K^-pp$ system, and 
$\langle \alpha \, | \psi_n ({\bm r}) | \, i \rangle$ 
is the $2N$-$N$ wave function for a struck neutron in the 
target where $\alpha$ denotes the complete set of eigenstates 
for the system. 
$\chi^{(+)}$ and $\chi^{(-)}$
are distorted waves of the incoming $K^-$ with the momentum 
${\bm p}_{K^-}$ and the outgoing $n$ with ${\bm p}_n$, 
respectively. 
The factors of $M_{\rm C}/M_{K^-pp}$ and $M_{\rm C}/M_{\rm A}$ 
in Eq.(\ref{eq:chi+-}) take into account the recoil effects, 
where $M_{\rm C}$ and $M_{\rm A}$ are the masses
of the ``$pp$'' core-nucleus and the $^3$He target,
respectively.
The recoil effects have to moderate a whole shape of the spectrum.
Indeed, if the recoil factors are omitted,
the cross section of a peak in the bound region is reduced by 
about 50\%; a yield in the quasi-free (QF) region grows up, 
and a QF peak is shifted upward to the higher-energy side and 
its width is broader.
Here we actually use the factor of $M_{\rm C}/M_{K^-pp}$ 
in not only $\chi^{(-)}$ but also $\chi^{(+)}$ for simplicity.
If we use an alternative factor of $M_{\rm C}/\bar{M}_{AK}$ where 
$\bar{M}_{AK}$ is the mean mass of $M_{\rm A}$ and $M_{K^-pp}$
instead of $M_{\rm C}/M_{K^-pp}$, 
we find that the cross section of the peak in the bound region 
is enhanced by less than 10\%.

By Green's function technique, 
the strength function $S(E)$ for the inclusive spectrum can 
be easily decomposed into two parts~\cite{Koi07,Mor94}; 
\begin{eqnarray}  
S(E) &=& S^{\rm con}(E) + S^{\rm esc}(E), 
\label{eq:Sdecomp}
\end{eqnarray}
where $S^{\rm con}(E)$ denotes the $K^-$ conversion 
processes including the decay modes of 
$[K^-pp] \to \pi + Y +N$ and $[K^-pp] \to Y +N$, 
which come from the one- and two-nucleon $K^-$ 
absorptions in Eqs.~(\ref{eq:decaymode1}) 
and (\ref{eq:decaymode2}), respectively; 
$S^{\rm esc}$(E) denotes the $K^-$ escape processes 
where the $K^-$ leaves from the core-nucleus as 
$[K^-pp] \to$ $K^-$ + ``$pp$'' above the $K^- +p+p$ 
threshold ($E > 0$). 
By an abbreviated notation for $G(E)$, $U^{\rm opt}(E)$ 
and $f$ instead of those in Eq.~(\ref{eq:S(E)}), we have 
%
\begin{eqnarray} 
S^{\rm con}(E) &=& - \, \frac{1}{\pi} 
\left\langle 
f^{\dagger}G^{\dagger}(E)\{ {\rm Im} \, U^{\rm opt}(E) \} G(E) f
\right\rangle,
\label{eq:Scon}
\\
S^{\rm esc}(E) &=& - \, \frac{1}{\pi}
\left\langle 
f^{\dagger}(1 +  G^{\dagger}(E)U^{\rm opt \dagger}(E))
\{ {\rm Im} \, G_0(E) \} (1 + U^{\rm opt}(E)G(E))f
\right\rangle,
\label{eq:Sesc}
\end{eqnarray}
where $G_0(E)$ is a free Green's function. 

With the help of the eikonal approximation, 
we express the distorted waves in Eq.~(\ref{eq:chi+-}), as
%
\begin{eqnarray}   
\chi^{(-)*} \left( {\bm p}_{n},
 {\bm r} \right)
&=&
\exp \left[ - i {\bm p}_{n}\cdot{\bm r}
- \frac{i}{v_n} 
\int^{+\infty}_{z} U_n({\bm b},z')dz'
\right],
\label{eq:EikonalWave-}
\\
\chi^{(+)} \left( {\bm p}_{K^-},
 {\bm r}  \right)
&=&
\exp \left[ + i {\bm p}_{K^-}\cdot {\bm r} 
- \frac{i}{v_{K^-}} 
  \int^z_{-\infty} U_{K^-}({\bm b},z')dz'
\right]
\label{eq:EikonalWave+}
\end{eqnarray}
with an impact parameter coordinate ${\bm b}$ and 
the optical potential for $\lambda=K^-$ or $n$,
\begin{eqnarray}   
U_{\lambda}(r) = -i \, {v_{\lambda} \over 2} \, 
\bar{\sigma}^{\rm tot}_{\lambda N} \, (1 - i \alpha_{\lambda N}) \, \rho(r), 
\label{eq:EikonalU}
\end{eqnarray}
where $\rho(r)$ is a nuclear density distribution, and 
$\bar{\sigma}^{\rm tot}_{\lambda N}$ and $\alpha_{\lambda N}$ 
denote the isospin-averaged total cross section
and the ratio of the real to imaginary parts of the forward amplitude
for the $\lambda + N$ scattering, respectively.
At $p_{K^-} =$ 1.0 GeV/c, 
the total cross sections of $\sigma^{\rm tot}_{K^- p}$ 
and $\sigma^{\rm tot}_{K^- n}$ amount to $\sim$50 mb and $\sim$40 mb, 
respectively, and $\sigma^{\rm tot}_{n p}$ varies 
within $30\text{--}40$ mb in the corresponding momenta 
$p_{n}=$ $1.1\text{--}1.4$ MeV/c~\cite{PDG}. 
We confirm that the absolute values of the formation cross section 
of Eq.~(\ref{eq:DWIA}) are enhanced or reduced by up to about 20\%
when the values of $\bar{\sigma}^{\rm tot}_{\lambda N}$ are changed 
within $30\text{--}50$ mb, but a whole shape of 
the spectrum is hardly moderated. 
Thus we use $\bar{\sigma}^{\rm tot}_{K^- N} 
= \bar{\sigma}^{\rm tot}_{nN} = 40$ mb~\cite{Kis99,Cie01} for simplicity.
Since the formation cross section is rather insensitive to 
the values of  $\alpha_{\lambda N}$~\cite{Dov80}, 
we use $\alpha_{K^-N}=\alpha_{nN}=0$~\cite{Kis99,Cie01}.
This fact implies that the distortion effects are not so 
important in our calculation, because of the small nuclear 
size of the $^3$He target. 
It has been shown that a distortion factor for the (in-flight $K^-$,~$n$) 
reactions on $^3$He is estimated as $D_{\rm dis}$= 0.47 for 
$1s_p \to 1s_{K^-}$ transition \cite{Koi07}, 
of which value is about 5 times as large as 
$D_{\rm dis}$=0.095 for $1p_p \to 1s_{K^-}$ transition 
on a $^{12}$C target \cite{Cie01}. 
This is also an advantage of a use of the $s$-shell nuclear 
targets such as $^3$He.

\subsection{Optical potentials for the $K^-$-``$\bm{pp}$'' system}
\label{Opt-pot}

\Figuretable{FIG. 1}

\begin{figure}[htb]
\begin{center}
  \includegraphics[angle=90,width=0.75\linewidth]{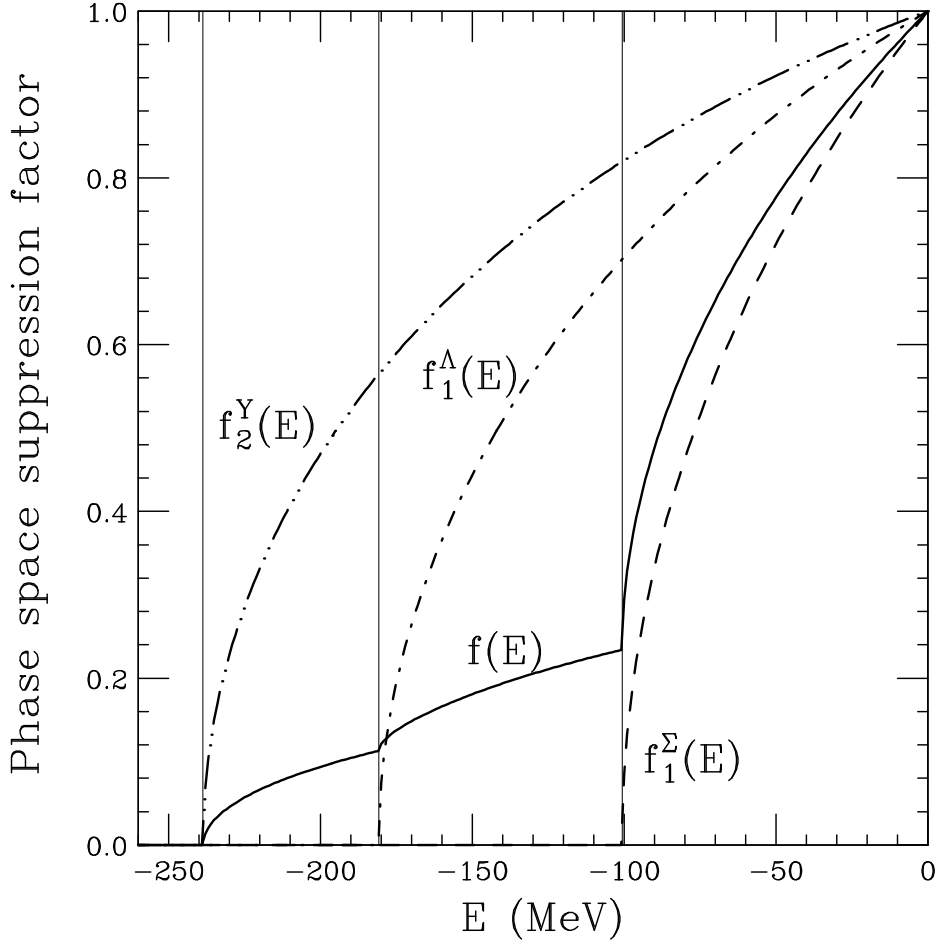}
  \caption{\label{fig:1} 
Phase space suppression factors introduced by 
Mare$\breve{\rm s}$ $et$ $al.$~\cite{Mar05},
as a function of $E$.
The dashed and dash-dotted curve denote
the phase space factors $f_1^{\Sigma}(E)$ 
for the $[K^-pp] \to \pi + \Sigma + N$ process
and $f_1^{\Lambda}(E)$ 
for the $[K^-pp] \to \pi + \Lambda + N$ one, respectively,
from the one-nucleon $K^-$ absorption.
The dash-dot-dotted curve denotes $f_2^Y(E)$
for the $[K^-pp] \to Y + N$ process from the 
two-nucleon $K^-$ absorption.
The solid curve denotes the total phase space factor 
$f(E) =  B_1^{(\pi \Sigma N)} f_1^{\Sigma}(E) + 
B_1^{(\pi \Lambda N)} f_1^{\Lambda}(E) + B_2^{(Y N)} f_2^Y(E)$,
where 
($B^{(\pi \Sigma N)}_1$,~$B^{(\pi \Lambda N)}_1$,~$B^{(Y N)}_2$) 
= (0.7,~0.1,~0.2) is assumed.
The vertical lines at $E \simeq -100$ MeV, $-180$ and $-240$ MeV 
indicate the $\pi + \Sigma + N$, $\pi + \Lambda + N$ and 
$Y + N$ decay threshold energies, respectively.
  }
  \end{center}
\end{figure}

\begin{figure}[htb]
  \begin{center}
  \includegraphics[angle=90,width=0.80\linewidth]{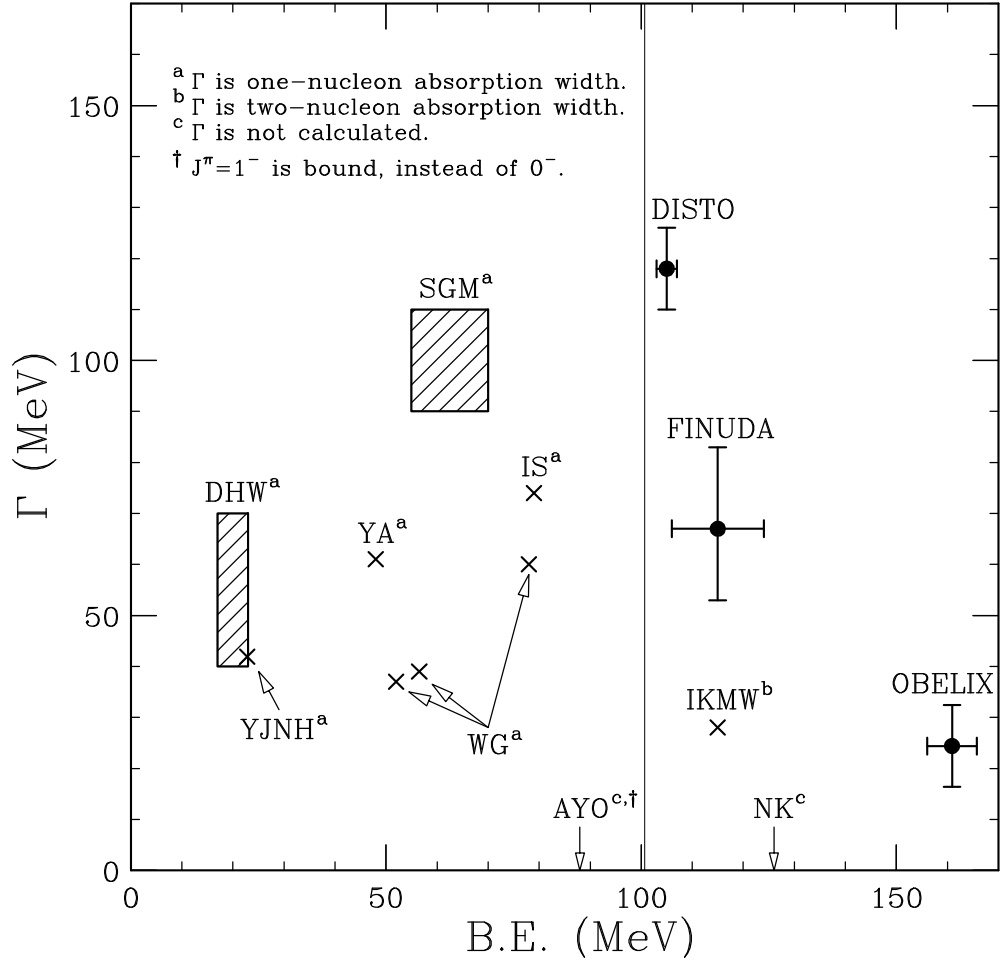}
  \caption{\label{fig:2}
  Summary of the binding energies and widths of the $K^-pp$ bound state. 
  Theoretical calculations with the $I=$ 1/2, $J^\pi= $ 0$^-$ state
  are predicted by
  YA (Yamazaki, Akaishi)~\cite{YA02}, 
  SGM (Shevchenko, Gal, Mare$\breve{\rm s}$)~\cite{She07}, 
  IS (Ikeda, Sato)~\cite{Ike07}, 
  DHW (Dot$\acute{\rm e}$, Hyodo, Weise)~\cite{Dot08},
  IKMW (Ivanov, Kienle, Marton, Widmann)~\cite{Iva05},
  NK (Nishikawa, Kondo)~\cite{Nis08},
  YJNH (Yamagata, Jido, Nagahiro, Hirenzaki)~\cite{Yam08},
  and WG (Wycech, Green)~\cite{Wyc08}; 
  the $I=$ 1/2, $J^\pi= $ 1$^-$ state by 
  AYO (Arai, Yasui, Oka)~\cite{Ara08}.
  The data are taken from the 
  FINUDA~\cite{FINUDA05}, OBELIX~\cite{OBELIX07} 
  and DISTO~\cite{DISTO08} experiments. 
  The vertical line at $B.E. \simeq 100$ MeV indicates
  the $\pi \Sigma N$ decay threshold.
}
  \end{center}
\end{figure}

In a previous paper~\cite{Koi07}, we evaluated the spectra 
with the energy-independent $K^-$-``$pp$'' optical potential which 
reproduces the result of the binding energy ($B.E.$) 
and width ($\varGamma$) by Yamazaki and Akaishi~\cite{YA02} 
or by Shevchenko, Gal and Mare$\breve{\rm s}$~\cite{She07}. 
On the other hand, Mare$\breve{\rm s}$ $et$ $al.$~\cite{Mar05} 
introduced phase space suppression factors, 
$f_1^Y(E)$ and $f_2^Y(E)$, 
which denote for the one- and two-nucleon $K^-$ absorption 
processes, respectively:
%
\begin{eqnarray}   
f_1^Y(E) &=&  
 \frac{M^3_1(0)}{M^3_1(E)}
\sqrt{\frac
{[M^2_1(E) - (M_{Y} + M_{\pi})^2 ]
 [M^2_1(E) - (M_{Y} - M_{\pi})^2 ]}
{[M^2_1(0) - (M_{Y} + M_{\pi})^2 ]
 [M^2_1(0) - (M_{Y} - M_{\pi})^2 ]}} \nonumber\\
& & \times {\mit\Theta}(M_1(E) - M_{Y}- M_{\pi} ), 
\label{eqn:f1}
\\
f_2^Y(E) &=&  
\frac{M^3_2(0)}{M^3_2(E)}
\sqrt{\frac{[ M^2_2(E) - (M_{Y} + M_{N})^2 ]
            [ M^2_2(E) - (M_{Y} - M_{N})^2 ]}
           {[ M^2_2(0) - (M_{Y} + M_{N})^2 ]
            [ M^2_2(0) - (M_{Y} - M_{N})^2 ]}}
\nonumber \\
&& \times {\mit\Theta}(M_2(E) - M_{Y} - M_{N})
\label{eqn:f2}
\end{eqnarray}
with $M_1(E)=M_{\bar{K}}+M_{N}+E$ and
$M_2(E)=M_{\bar{K}} + 2 \, M_{N} + E$,  
where $M_{\bar{K}}$, $M_{N}$, $M_Y$ and $M_{\pi}$ denote
the masses of $\bar{K}$, nucleon, hyperon ($Y = \Sigma$ or $\Lambda$), 
and $\pi$, respectively;
${\mit\Theta}(x) = 1$ for $x \geq 0$ and 0 for $x < 0$.
Fig.~\ref{fig:1} displays these phase space suppression factors, 
as a function of $E$. 
$f_1^{\Sigma}(E)$ vanishes below 
the $\pi \Sigma N$ threshold of $E_{\rm th}(\pi \Sigma N)=-101$ MeV,
and $f_1^{\Lambda}(E)$ vanishes below 
the $\pi \Lambda N$ threshold of $E_{\rm th}(\pi \Lambda N)=-181$ MeV. 
$f_2^{Y}(E)$ vanishes below the $Y N$ threshold, e.g., 
$E_{\rm th}(\Sigma N)=-239$ MeV or $E_{\rm th}(\Lambda N)=-319$ MeV. 
As attempted in Refs.~\cite{Kis05,Yam06}, we take into account
the energy dependence of the imaginary part
multiplied by $f_1^Y(E)$ or $f_2^Y(E)$ in the optical potential. 
Thus we employ the energy-dependent $K^-$-``$pp$'' optical potential
which is parametrized in a Gaussian form with a range parameter $b$, 
as
%
\begin{equation}
U^{\rm opt}(E; {\bm r}) = 
\left( V_0 + i \, W_0 f(E) \,  \right) \exp[-({\bm r}/b)^2]
\label{eq:E-dep_opt}  
\end{equation}
with 
%
\begin{equation}
f(E) = B^{(\pi \Sigma N)}_1 f_1^{\Sigma}(E) +
       B^{(\pi \Lambda N)}_1 f_1^{\Lambda}(E) +
       B^{(Y N)}_2  f_2^Y(E),
\label{eq:totf}
\end{equation}
where $V_0$ and $W_0$ are adjusted parameters, 
of which values are determined to 
reproduce the result of the binding energy and width of 
the $K^-pp$ state in theoretical predictions or 
experimental data, as we will mention below. 
$B^{(\pi \Sigma N)}_1$ and $B^{(\pi \Lambda N)}_1$
are branching rates to $[K^-pp] \to \pi + \Sigma + N$ 
and $[K^-pp] \to \pi + \Lambda+ N$ 
decay channels from the one-nucleon $K^-$ absorption, respectively, 
and $B^{(Y N)}_2$ is a branching rate to $[K^-pp] \to Y + N$
decay channel from the two-nucleon $K^-$ absorption. 
Here we assumed~\cite{Mar05,Gaz07,Yam06}
\begin{equation}
B^{(\pi \Sigma N)}_1 = 0.7,\quad
B^{(\pi \Lambda N)}_1 = 0.1, \quad
B^{(Y N)}_2= 0.2, 
\label{eq:rate}
\end{equation}
where we treat that the $[K^-pp] \to Y +N$ process acts effectively 
in $\Sigma +N$ and $\Lambda +N$ decay channels
because these channels similarly affect the spectrum 
within the present framework (see also Fig.~\ref{fig:1}).
Then we can rewrite the imaginary part of $U^{\rm opt}(E)$ 
in Eq.~(\ref{eq:E-dep_opt}) as 
%
\begin{eqnarray} 
{\rm Im } \ U^{\rm opt}(E; {\bm r})
= W_1^{\Sigma}(E; {\bm r}) + W_1^{\Lambda}(E; {\bm r}) + W_2^Y(E; {\bm r}),
\label{eq:ImU}
\end{eqnarray} 
where $W_1^Y(E;{\bm r})$ and $W_2^Y(E;{\bm r})$ correspond to 
the absorptive potentials for one- and two-nucleon $K^-$ absorptions, 
respectively: 
%
\begin{eqnarray} 
W_1^Y(E; {\bm r}) &=& B_1^{(\pi Y N)} W_0 \, f_1^Y(E) \, \exp[-({\bm r}/b)^2],
\label{eq:W1} \\
W_2^Y(E; {\bm r}) &=& B_2^{(Y N)}\, W_0 \, f_2^Y(E) \,  \exp[-({\bm r}/b)^2]. 
\label{eq:W12}
\end{eqnarray} 
In the Green's function method, 
$S^{\rm con}(E)$ of Eq.~(\ref{eq:Scon}) can be further 
decomposed~\cite{Koi07} as
\begin{eqnarray}  
S^{\rm con}(E) &=& S_{\pi\Sigma N}^{\rm con}(E) 
                +  S_{\pi\Lambda N}^{\rm con}(E) 
                +  S_{YN}^{\rm con}(E)
\label{eq:S12}
\end{eqnarray}
with 
\begin{eqnarray} 
S_{\pi Y N}^{{\rm con}}(E)  &=& - \, \frac{1}{\pi} 
\left\langle
f^{\dagger}G^{\dagger}(E) W_1^Y(E) G(E) f
\right\rangle,
\label{eq:S1Y}
\\
S_{YN}^{{\rm con}}(E) &=& - \, \frac{1}{\pi} 
\left\langle
f^{\dagger}G^{\dagger}(E)W_2^Y(E) G(E) f
\right\rangle, 
\label{eq:S2}
\end{eqnarray}
where $S_{\pi Y N}^{{\rm con}}(E)$ and $S_{YN}^{{\rm con}}(E)$ 
express components of the strength functions for 
the $\pi+ Y + N$ decay process from the one-nucleon $K^-$ absorption 
and for the $Y + N$ decay one from the two-nucleon $K^-$ absorption,
respectively, in the $K^-$ conversion spectra. 
Therefore, the semi-exclusive spectra in the $^3$He(in-flight $K^-$, $n$) 
reaction,
\begin{eqnarray} 
K^- + {^3}{\rm He} \to n + [K^-pp] \to n + Y +X,
\label{eq:semi-ex}
\end{eqnarray}
can be evaluated in our calculations, 
where $Y=\{\Sigma, \Lambda\}$ and $X=\{\pi+N, N\}$.

\Figuretable{FIG. 2}

The binding energies $B.E.$ and widths ${\varGamma}$ 
of the $K^-pp$ bound state with $I=1/2$, $J^\pi=0^-$ 
have been predicted in many 
calculations~\cite{She07,Ike07,Dot08,Iva05,Nis08,Ara08,Yam08,Wyc08} 
and also reported in several 
experiments~\cite{FINUDA05,OBELIX07,DISTO08}.
In Fig.~\ref{fig:2}, we summarize the values of $B.E.$ and ${\varGamma}$
taken from theoretical predictions and experimental candidates.
By considering these results of $B.E.$ and ${\varGamma}$ as a guide, 
we attempt to construct the $K^-$-``$pp$'' optical potentials $U^{\rm opt}(E)$. 
We solve the Klein-Gordon equation self-consistently 
in the complex energy plane:
%
\begin{eqnarray}   
\{ ( \omega(E) - V_{\rm Coul}({\bm r}) )^2  
 + {\bm \nabla}^2 - \mu^2 - 2 \, \mu \, 
   U^{\rm opt}(E; {\bm r}) \}
\, \varPhi(E; {\bm r}) 
&=& \, 0, 
\label{eq:KG}
\end{eqnarray}
where $\varPhi(E; {\bm r})$ is a relative wave function between 
the $K^-$ and ``$pp$" core nucleus, and $\omega(E)$ is a complex 
eigenvalue, as a function of $E$ which is a real number. 
If we find that $E$ satisfies 
%
\begin{equation}   
{\rm Re} \ \omega(E) = E,
\label{eq:defBE}
\end{equation}
we can obtain  ${\rm Re} \ \omega(E) = - B.E.$ and 
${\rm Im} \ \omega(E) = -{\varGamma}/2$ as the Klein-Gordon complex 
energy.
Thus we determine the strength parameters of ($V_0$, $W_0$) 
in Eq.~(\ref{eq:E-dep_opt}) by fitting to 
the prediction or candidate of $B.E.$ and ${\varGamma}$. 
Here we introduce four $K^-$-``$pp$'' optical potentials 
$U^{\rm opt}(E)$ as follows:

\begin{itemize}

\item[(a)] 
potential A which we determined 
by fitting to $B.E. \simeq$ 20 MeV and 
the maximum ${\varGamma} \simeq$ 70 MeV in 
a variational three-body calculation based on the 
Chiral unitary approach 
by Dot$\acute{\rm e}$, Hyodo and Weise~\cite{Dot08};

\item[(b)] 
potential B which is equivalent to 
the energy-independent optical potential obtained in a variational 
three-body calculation by Yamazaki and Akaishi~\cite{YA02}, 
by fitting to $B.E. =$ 48 MeV and ${\varGamma}=$ 61 MeV;

\item[(c)] 
potential C which we determined by fitting 
to $B.E. \simeq$ 70 MeV and ${\varGamma} \simeq$ 110 MeV
in a $\bar{K}NN$-$\pi \Sigma N$ coupled-channel Faddeev calculation
by Shevchenko, Gal and Mare$\breve{\rm s}$~\cite{She07}.
These values correspond to the maximum $B.E.$ and ${\varGamma}$, 
respectively, within the uncertainty of their results; 

\item[(d)] 
potential D which is a series of the potentials we determined by 
fitting to the experimental observations of $B.E.$ and ${\varGamma}$: 
D$_1$, D$_2$ and D$_3$ indicate the potentials 
for DISTO~\cite{DISTO08}, FINUDA~\cite{FINUDA05} and 
OBELIX~\cite{OBELIX07} experiments, respectively.  

\end{itemize}
For the range parameter for $U^{\rm opt}(E)$ 
in Eq.~(\ref{eq:E-dep_opt}), here we used $b= 1.09$ fm, 
of which value is derived from the results of three-body 
calculations by Yamazaki and Akaishi~\cite{YA02}. 
A dependence of the spectrum on the range parameter $b$ 
is slightly seen in the QF region;
for example, when $b$ is changed within $+ 0.12$ fm ($- 0.12$ fm) 
in potential C, the cross section of a QF peak is 
reduced (enhanced) by less than 10\% and its peak position is shifted 
within $- 5$ MeV ($+ 5$ MeV), 
while the bound-state spectrum is almost unchanged.
Note that the values of ${\varGamma}$ 
which we considered in (a)-(c), are obtained by microscopic 
three-body calculations
with only $\pi + Y +N$ decay processes in the 
one-nucleon $K^-$ absorption~\cite{She07,YA02,Dot08}.
Since a $K^-$-``$pp$'' optical potential has to describe 
not only one-nucleon $K^-$ absorption processes 
but also two-nucleon ones, 
we employ the parameters of $(V_0,~W_0)$ 
for  potentials A, B and C by fitting to the values of 
$B.E.$ and ${\varGamma}$ without $B_2^{(Y N)}$ in 
Eq.(\ref{eq:totf}), i.e., 
($B_1^{(\pi \Sigma N)}$,~$B_1^{(\pi \Lambda N)}$,~$B_2^{(Y N)}$)
=(0.7,~0.1,~0.0).
For potentials D, we took the parameters with $B_2^{(Y N)}$. 
In Table~\ref{tab:1}, we list the parameter sets of ($V_0$, $W_0$) for 
$U^{\rm opt}(E)$, 
together with their calculated binding energies $B.E$ and widths ${\varGamma}$ 
of the $K^-pp$ bound state.
We find that when $B^{(YN)}_2$ is switched on, the 
values of $B.E.$ decrease and those of ${\varGamma}$ increase.

\begin{figure}[htb]
  \begin{center}
  \includegraphics[angle=90,width=0.98\linewidth]{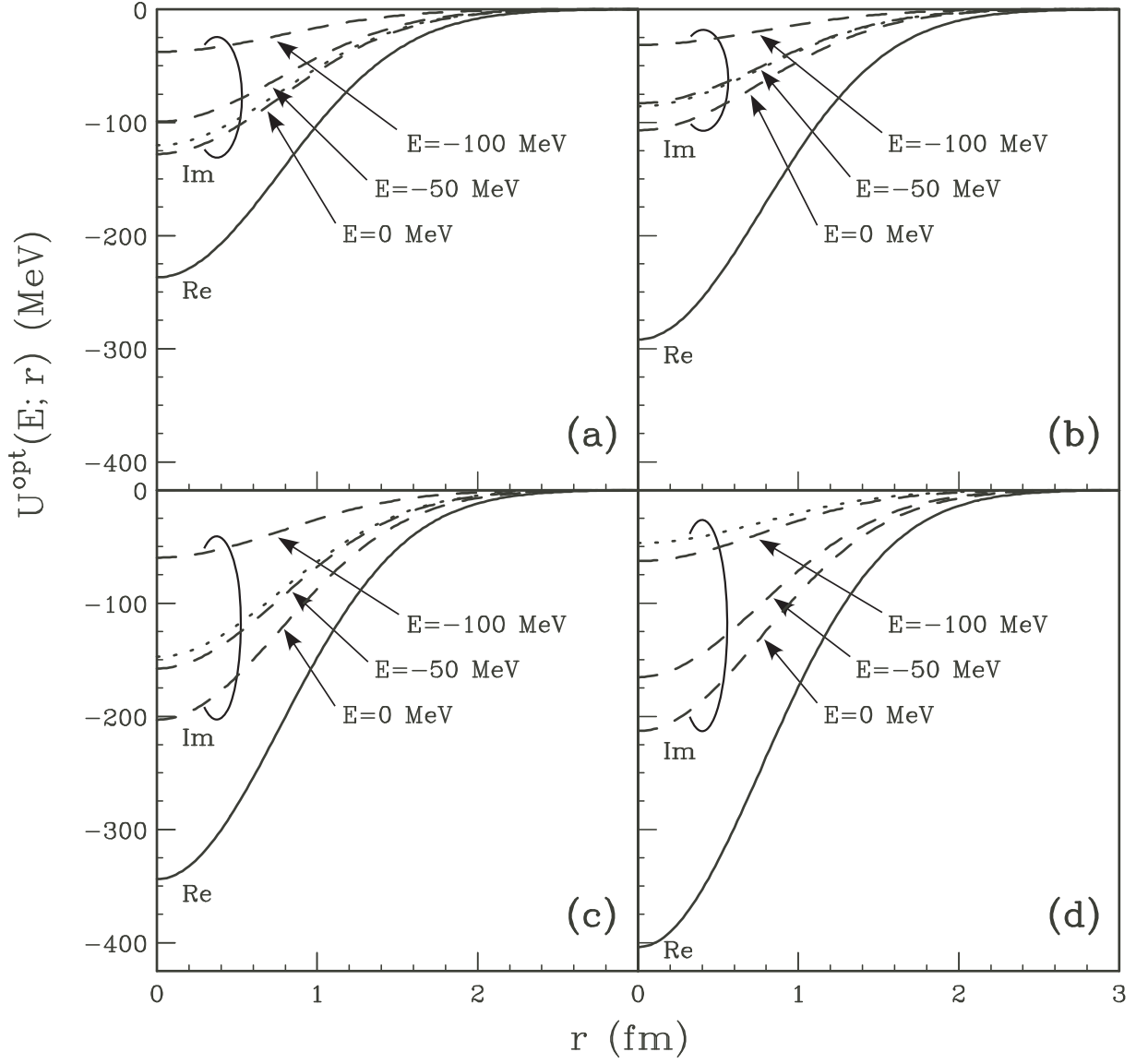}
  \caption{\label{fig:3}
  Real and imaginary parts of the $K^-$-``$pp$" optical potentials 
  $U^{\rm opt}(E;{\bm r})$ for potentials 
  (a) A,
  (b) B,
  (c) C and
  (d) D$_2$,
  as a function of a distance between 
  the $K^-$ and the center of the ``$pp$" core nucleus. 
  The solid curves denote the real parts, and 
  the dashed curves the imaginary parts at $E = 0, -50$ and $-100$ MeV.
  The dotted curves denote the imaginary parts for 
  the energy-independent $K^-$-``$pp$'' optical potentials
  $U^{\rm opt}_0({\bm r})$.
}
  \end{center}
\end{figure}

In Fig.~\ref{fig:3}, we display the real and imaginary parts of the 
$K^-$-``$pp$'' optical potentials $U^{\rm opt}(E)$, as a function of 
a distance between the $K^-$ and the center of the ``$pp$'' core nucleus. 
If we neglect the energy-dependence in $U^{\rm opt}(E)$ of 
Eq.~(\ref{eq:E-dep_opt}) by replacing $f(E)$ by 1, 
we find the energy-independent optical potentials $U^{\rm opt}_0$, 
as used in our previous calculation~\cite{Koi07}.
It should be noticed the values of $W_0$ in $U^{\rm opt}_0$ differ from 
those in $U^{\rm opt}(E)$, as shown in Table~\ref{tab:1},
whereas we have 
${\rm Im} \, U^{\rm opt}_0={\rm Im} \, U^{\rm opt}(E)$ 
at $E = - B.E.$

\Figuretable{TABLE I}

\Figuretable{FIG. 3}

\section{Numerical Results}
\label{Results}

\subsection{Inclusive spectrum by the $^3$He(in-flight $K^-$,~$n$) reaction}
\label{inclusive}

Let us consider the $^3$He(in-flight $K^-$,~$n$) reaction
at $p_{K^-} = 1.0$ GeV/c and $\theta_{\rm lab}=$ 0$^\circ$ 
for the J-PARC E15 experiment \cite{Iwa06}.
To find possible evidence of the $K^-pp$ bound state, 
we evaluate the inclusive and semi-exclusive spectra of the 
$^3$He(in-flight $K^-$,~$n$) reaction numerically
by Eqs.~(\ref{eq:DWIA}), (\ref{eq:Sdecomp})-(\ref{eq:Sesc})
and (\ref{eq:S12})-(\ref{eq:S2}).

\Figuretable{FIG. 4}

\begin{figure}[htb]
\begin{center}
  \includegraphics[angle=90,width=\linewidth]{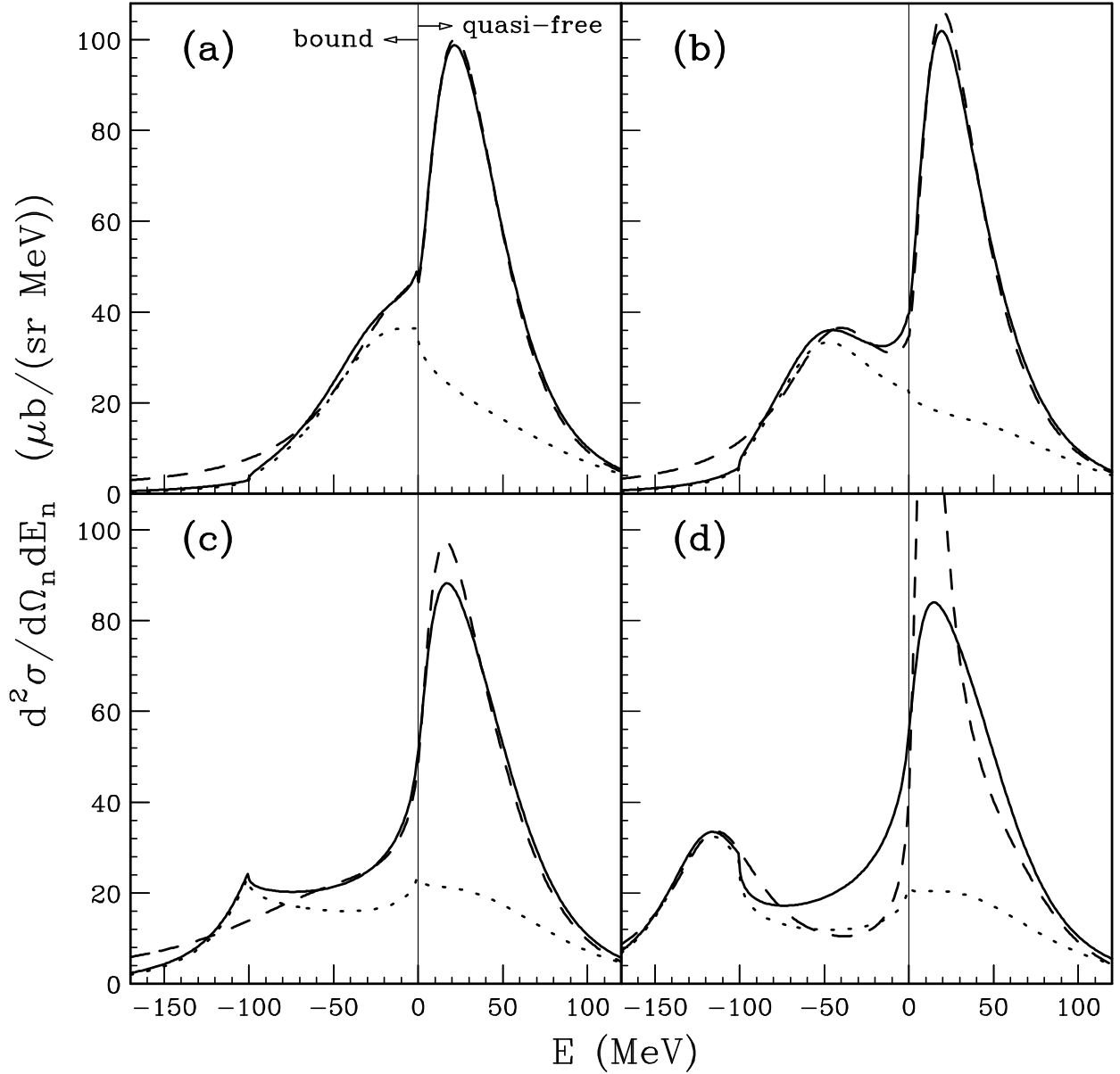}
  \caption{\label{fig:4} 
Calculated inclusive spectra of 
the $^3$He(in-flight $K^-$, $n$) reaction at 
$p_{K^-} = 1.0$ GeV/c and $\theta_{\rm lab}=$ 0$^\circ$
as a function of the energy $E$ of the $K^- pp$ system 
measured from $K^-+p+p$ threshold for potentials 
  (a) A,  (b) B,  (c) C and  (d) D$_2$.
The solid and dashed curves denote the inclusive spectra with 
the energy-dependent $U^{\rm opt}(E)$ and energy-independent 
$U^{\rm opt}_0$ potentials, respectively. 
The dotted curve denotes the $L=0$ component in the inclusive 
spectrum for $U^{\rm opt}(E)$. 
The vertical line at $E = 0$ MeV 
indicates the $K^- + p + p$ threshold, and the left- and right-hand 
sides of this line are the $K^-$ bound and quasi-free scattering 
regions, respectively. 
  }
  \end{center}
\end{figure}

In Fig.~\ref{fig:4}, we display the calculated results of the 
inclusive spectra with the optical potentials $U^{\rm opt}(E)$ 
listed in Table~\ref{tab:1}.
In Fig.~\ref{fig:4}(c), we show the calculated inclusive spectrum 
for potential C
where the binding energy and width of the $K^-pp$ bound state 
are obtained as $B.E. = 59$ MeV and ${\varGamma} = 164$ MeV, 
respectively. The inclusive spectrum with $U^{\rm opt}(E)$
is qualitatively different from that with the energy-independent 
$U^{\rm opt}_0$. 
The former has a cusp which appears at the $\pi\Sigma N$ threshold 
in the $L=0$ component of the spectrum, 
whereas the latter has no peak due to its large width of 
the $K^-pp$ state~\cite{Koi07}.
Such a cusp-like structure originates from the energy dependence of the 
imaginary part of $U^{\rm opt}(E)$, and its mechanism can be understood 
by behavior of a pole trajectory of the $K^-pp$ state 
in the complex energy plane, as we will discuss in 
Sect.~\ref{Mechanism_Cusp}.

In Fig.~\ref{fig:4}(b), we show the inclusive spectrum 
for potential B
which gives $B.E. = 45$ MeV and ${\varGamma} = 82$ MeV.
A clear peak of the $K^-pp$ state appears 
in both the spectra with $U^{\rm opt}(E)$ and $U^{\rm opt}_0$,
but the peak position for $U^{\rm opt}(E)$ is slightly shifted
from $E=-45$ MeV to $-50$ MeV because of its energy dependence.
Since this state is away from 
the branching point of the $\pi \Sigma N$ threshold, 
the peak in the spectrum is scarcely influenced by the threshold, 
so that its shape approximately indicates a standard Breit-Wigner 
(BW) form~\cite{Koi08}. 
Since a $\pi \Sigma N$ phase space is taken into account, 
the spectrum for $U^{\rm opt}(E)$ is suppressed below the 
$\pi \Sigma N$ decay threshold of 
$E_{\rm th}(\pi \Sigma N) \simeq -100$ MeV, 
in contrast to that for $U^{\rm opt}_0$~\cite{Koi07}.

In Fig.~\ref{fig:4}(a), 
we find that a peak of the $K^-pp$ bound state 
is not clear in the inclusive spectrum with potential A,
because of the relatively small binding energy of 15 MeV 
with the large width of 92 MeV. 
The energy dependence of the spectrum for $U^{\rm opt}(E)$ 
seems to be similar to that of potential B.
Yamagata $et$ $al.$~\cite{Yam08} also performed 
a similar calculation using the energy-dependent $K^-$-``$pp$'' 
optical potential based on the Chiral unitary model.
The shape of their inclusive spectrum is 
different from that of ours 
because of a different contribution of the partial-wave components
and their different widths; 
they found a $K^-pp$ $L=0$ bound state with 
$B.E. \simeq 20$ MeV and ${\varGamma} \simeq 40$ MeV, and 
other $L=$ 1, 2 bound states with $B.E. \simeq 10$ MeV~\cite{Yam08}. 
However, it is not understood that the $K^-pp$ system has 
nuclear bound states with $L \geq 1$, 
in our consideration to its small nuclear size such as $^3$He~\cite{NoteAdd}.

We consider potential D$_2$ as a typical example of the potential
D series. In Fig.~\ref{fig:4}(d),
we show the inclusive spectrum 
with potential D$_2$ 
by fits to the experimental values, 
$B.E. = 115 $ MeV and ${\varGamma} = 67$ MeV,
which are obtained from the invariant-mass spectrum 
by the FINUDA experiment~\cite{FINUDA05}.
If an interpretation of the FINIDA candidate as a $K^-pp$ bound state 
is true, a clear peak should appear below the $\pi \Sigma N$ threshold
in the missing-mass spectrum.
It would be easy to observe such a peak structure experimentally.
The spectra with $U^{\rm opt}(E)$ and $U^{\rm opt}_0$
are quite similar in shape below the $\pi \Sigma N$ threshold, 
whereas they are considerably different each other above the $\pi \Sigma N$ 
threshold. 
For potentials D$_1$ and D$_3$ which are determined respectively 
by fits to the data of DISTO~\cite{DISTO08} and OBELIX~\cite{OBELIX07}, 
we also obtain that a clear peak in their spectra 
appears below the $\pi \Sigma N$ threshold (see also Fig.~\ref{fig:10}). 
We find that the shape of the peak in the spectrum for D$_1$ is similar to 
that for D$_2$, as shown in Fig.~\ref{fig:4}(d), 
whereas the spectrum for D$_3$ has a very prominent peak 
because of the small width of $\varGamma = 24$ MeV.

Therefore, we recognize that the inclusive spectrum of the
$^3$He(in-flight $K^-$, $n$) reaction at $p_{K^-}=$ 1.0 GeV/c 
and $\theta_{\rm lab}=$ 0$^\circ$
is expected to have a signal for clear evidence of the $K^-pp$ 
bound state.
The calculated spectrum predominately has a bound state with 
$I=$ 1/2, $J^\pi=$ 0$^-$ and an orbital angular momentum 
$L = 0$ in the $K^-pp$ bound region below the $K^- +p+p$ threshold, 
whereas continuum states with $L \geq 1$ occur in the QF region.
We stress that the (in-flight $K^-$,~$n$) reaction 
on the $s$-shell nuclear target such as $^3$He 
provides an advantage to produce the deeply-bound 
$K^-$ nuclear state with $L=0$. 
These results also indicate the importance of the energy-dependence
in the $K^-$-``$pp$'' optical potentials, particularly, 
in the case of potentials C, D$_1$ and D$_2$. 
We realize the limit to apply the energy-independent optical 
potential to calculations for the spectrum; 
such a potential can be justified only for potential A   
involving the shallow potentials~\cite{Dot08,Yam08}, and it 
works approximately for potential B around the $K^-pp$ peak.
The strength of $|V_0|$ in $U^{\rm opt}_0$ must be shallower
than 300 MeV, which corresponds to $B.E. < 50$ MeV.

\subsection{Contribution of one- and two-nucleon $K^-$ absorption processes
in the $K^-$ conversion spectra}
\label{semi-exclusive}

\Figuretable{FIG. 5}

It should be noticed that the calculated $K^-$ conversion spectra
can be directly compared with the 
experimental data at the J-PARC E15 experiment, which is planned 
to measure the (in-flight $K^-$,~$n$) spectra and the decaying 
particles from the $K^- pp$ system simultaneously~\cite{Iwa06}; 
no $K^-$ escape spectrum will be measured in this experiment.
To search a better signal for the $K^-pp$ bound state, 
we focus on the $K^-$ conversion spectra which 
express semi-exclusive $K^- + {^3{\rm He}} \to n + Y + X$ 
processes in the deeply-bound region,
where $Y=\{\Sigma, \Lambda\}$ and $X=\{\pi+N, N\}$.
By Eqs.~(\ref{eq:S1Y}) and~(\ref{eq:S2}), 
we can calculate their strength functions 
for $K^-$ conversion processes, which are effectively described 
by the imaginary potential, 
${\rm Im} \, {U^{\rm opt}}(E)$ in Eq.~(\ref{eq:ImU}), 
within the optical models. 
For the $K^-$ escape processes, 
we can calculate the strength function in Eq.~(\ref{eq:Sesc}), 
which is probably underestimated because it has be 
described as $K^-$+``$pp$'' continuum states 
above the $K^-$ + $pp$ threshold 
where ``$pp$'' should actually break up into $p + p$.

\begin{figure}[htb]
\begin{center}
\includegraphics[angle=90,width=\linewidth]{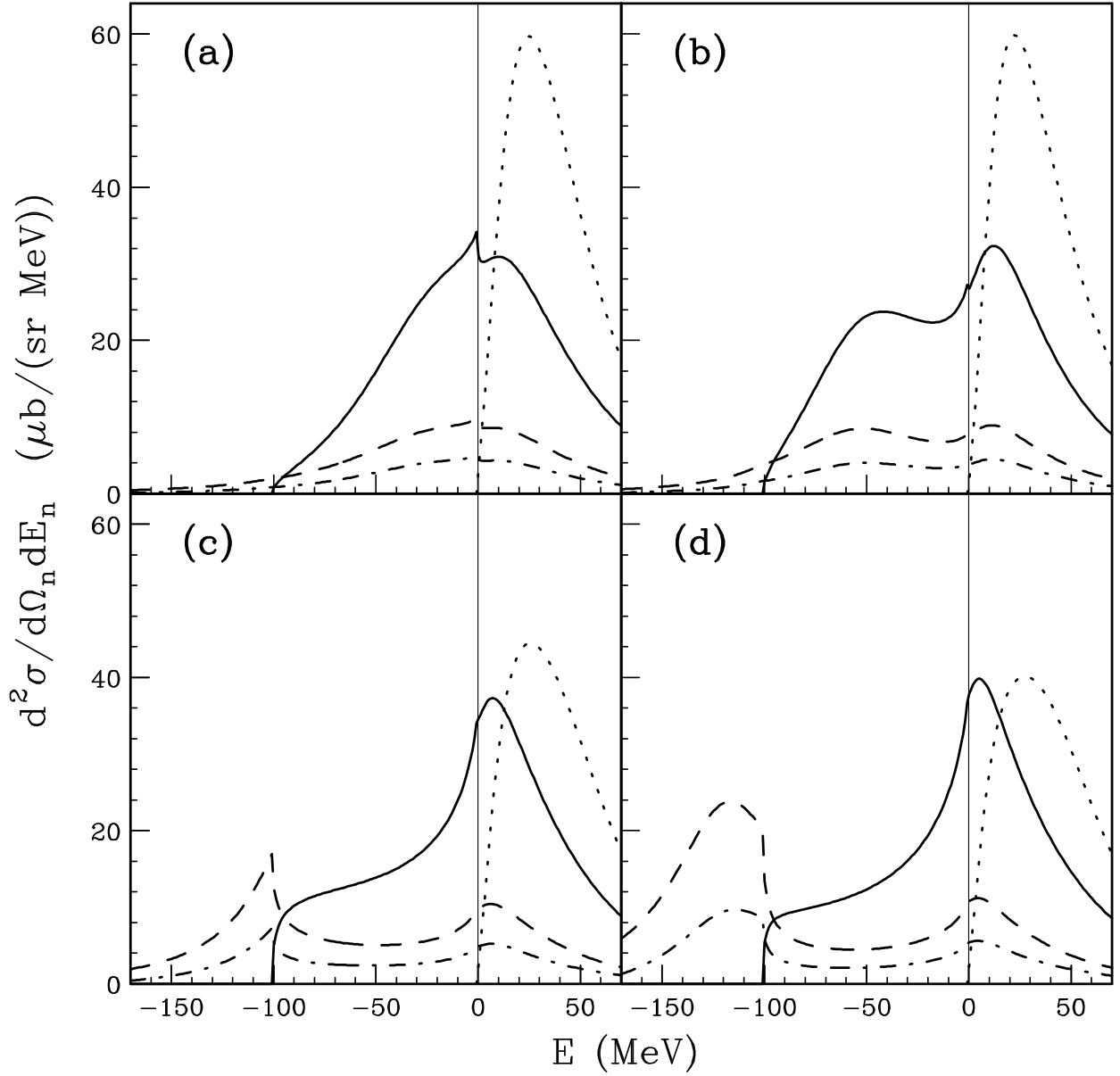}
\caption{\label{fig:5}
Calculated semi-exclusive spectra
of the $^3$He(in-flight $K^-$, $n$) reaction at 
$p_{K^-} = 1.0$ GeV/c and $\theta_{\rm lab}=$ 0$^\circ$,
for potentials 
  (a) A,  (b) B,  (c) C and  (d) D$_2$.
The solid and dot-dashed curves denote 
the $[K^-pp] \to \pi + \Sigma + N$ and $\pi + \Lambda + N$
decay processes from the one-nucleon $K^-$ absorption, 
respectively.
The dashed curves denote the 
$[K^- pp] \to Y + N$ decay process from the 
two-nucleon $K^-$ absorption.
The dotted curves denote the spectra of the $K^-$ escape process. 
See also the caption in Fig.~\ref{fig:4}.
}
  \end{center}
\end{figure}

In Fig.~\ref{fig:5}, we illustrate partial contributions of 
the $^3$He(in-flight $K^-$,~$n$) spectrum
at $p_{K^-}=$ 1.0 GeV/c and $\theta_{\rm lab}=$ 0$^\circ$, e.g.,
$[K^-pp] \to \pi +\Sigma + N$ decay process 
and $[K^-pp] \to \pi + \Lambda + N$ decay process
from the one-nucleon $K^-$ absorption,
and $[K^-pp] \to Y + N$ decay process from the two-nucleon $K^-$ 
absorption. 
In Fig.~\ref{fig:5}(c), we find that a cusp-like peak 
at $\pi \Sigma N$ threshold originates from 
the $\pi+ \Lambda+ N$ and $Y+N$ decay channels in 
the spectrum for potential C, 
rather than the $\pi + \Sigma + N$ decay one
which is suppressed below the $\pi \Sigma N$ threshold because
its decay channel is closed there.
Such a cusp-like peak is clear evidence of the 
formation of the $K^-pp$ bound state. 
Since the observed peak position and width do not directly
correspond to the actual binding energy and width, respectively, 
we need a comparison between the theoretical 
and experimental spectra to extract the binding energy 
and width of the $K^-pp$ bound state 
from the cusp-like peak. 

In Fig.~\ref{fig:5}(d), 
it is also interesting to see a clear peak in the 
$\pi + \Lambda + N$ and $Y + N$ decay spectra
for potential D$_2$
which is more attractive than potential C. 
Its shape is asymmetric because it must be sharply cut 
by the phase space suppression factor
above the $\pi \Sigma N$ threshold.
Moreover, we confirm that the $\pi + \Sigma + N$  decay spectrum gives 
no peak for the $K^-pp$ bound state, as well as 
the spectrum for potential C.

In Fig.~\ref{fig:5}(b), we show partial contributions 
of the calculated spectrum with potential B.
We find that the shape of the spectra with $\pi + \Sigma + N$, 
$\pi + \Lambda  + N$ and $Y + N$
decay processes are essentially the same,
but the peak positions of their spectra are slightly different 
each other because of the energy dependence of the potential.
A clear signal would be observed in the spectrum with 
$[K^-pp] \to Y + N$ from the two-nucleon $K^-$ absorption,
as well as the inclusive spectrum shown in Fig.~\ref{fig:4}(b). 

On the other hand, we confirm that 
there is no peak in any partial contributions 
with potential A 
even if the $K^-pp$ bound state exists, 
as shown in Fig.~\ref{fig:5}(a). 
This state exists close to the $K^-+p+p$ threshold due to a 
small binding energy of $B.E. = 15$ MeV 
and a large width of $\varGamma =$ 92 MeV.
For a more quantitative estimation, we need to examine a whole 
shape of the spectrum including the effects of the $K^- +p+p$ 
threshold~\cite{Mor94,Koi08}.

We recognize that the detailed comparison between the theoretical 
and experimental spectra is required to extract the binding energy 
and width of the $K^-pp$ bound state from the spectra.
The shape behavior of the $[K^-pp] \to \pi + \Lambda + N$ decay 
spectrum is quite similar to that of the $[K^-pp] \to Y + N$ one 
in all of our potentials.
This similarity is understood from the fact that 
the energy dependence of the phase space factor 
$f_1^\Lambda(E)$ for the $\pi + \Lambda + N$ decay processes
resembles that of $f_2^Y(E)$ 
for the $Y + N$ decay processes near the $\pi \Sigma N$ threshold.

\subsection{Pole trajectory for the deeply-bound $K^-pp$ state}
\label{Mechanism_Cusp}

\subsubsection{Moving pole in the complex energy plane}

\Figuretable{FIG. 6}

It is important to understand the mechanism of a peak 
structure near the $\pi \Sigma N$ threshold 
in the spectrum, so as to identify the nature of 
the $K^-pp$ bound state from the experimental data. 
Quantum mechanically, the peak structure in the energy spectrum 
is associated with a pole in the scattering amplitude or 
the complete Green's function. The pole position corresponds 
to a complex eigenvalue of a Hamiltonian on the complex energy plane.
The shape of the spectrum must be modified by the threshold effects
if the pole is located near the branch point of the threshold. 
To understand the shape behavior of the $[K^-pp] \to Y + N$ 
decay spectrum, 
we investigate the pole position of the $K^-pp$ state 
in the complex energy plane.
We can obtain the pole position as a complex eigenvalue 
of $\omega(E)$ in Eq.~(\ref{eq:KG}) as a function of $E$
because of the energy dependence of $U^{\rm opt}(E)$. 

\begin{figure}
  \begin{center}
\includegraphics[angle=90,width=0.80\linewidth]{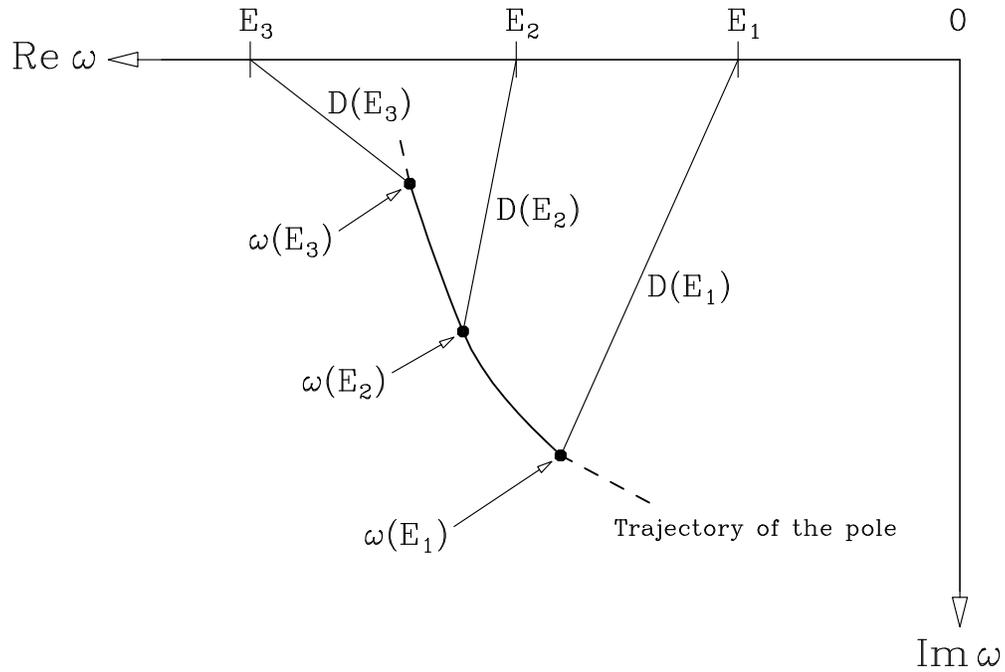}
\caption{\label{fig:6}
Distance $D(E)$ between a point ($E$, 0) on the real axis and 
a pole at (${\rm Re}\, \omega(E)$,~${\rm Im}\, \omega(E)$) 
for the $K^-pp$ state in the complex energy plane. 
The pole moves on the trajectory, as a function of the real 
energy $E$. Some examples are illustrated 
at $E=E_i$ ($i =$ 1, 2, 3).
  }
  \end{center}
\end{figure}

The shape of the inclusive spectrum in the $K^-$ bound 
region is perhaps written as the following form:
\begin{eqnarray}
S^{\rm (pole)}(E) &=& - \frac{1}{\pi} \, 
\frac{{\rm Im} \, \omega(E)}{D^2(E)},
\label{eq:SBW}
\end{eqnarray}
where 
%
\begin{eqnarray}   
D(E) &\equiv& 
\sqrt{(E-{\rm Re} \, \omega(E))^2 + ({\rm Im} \, \omega(E))^2 }
\label{eq:D(E)}
\end{eqnarray}
denotes a distance between a point ($E$,~0) of the physical state 
on the real axis and 
the pole at a point (${\rm Re}\, \omega(E)$,~${\rm Im}\, \omega(E)$) in 
the complex energy plane, as illustrated in Fig.~\ref{fig:6}.
If the energy dependence of $\omega(E)$ is negligible, 
the shape of $S^{\rm (pole)}(E)$ is equivalent to the BW resonance form. 
Since ${\rm Im} \, \omega(E)$ is approximately 
proportional to the phase space suppression factor $f(E)$,
the shape of the inclusive spectrum is roughly denoted by 
$f(E)/D^2(E)$.
Functions of $f_1^Y(E)/D^2(E)$ and $f_2^Y(E)/D^2(E)$ 
can simulate the shapes of the one- and two-nucleon $K^-$
absorption spectra, respectively.
It is apparent that $S_{\pi\Sigma N}^{{\rm con}}(E)$ 
is suppressed below the $\pi \Sigma N$ threshold
due to the behavior of the function $f_1^Y(E)$.
On the other hand, 
$S_{\pi\Lambda N}^{{\rm con}}(E)$ and $S_{YN}^{{\rm con}}(E)$ 
are approximately equivalent to $1/D^2(E)$ 
because $f_1^{\Lambda}(E)$ and $f_2^Y(E)$ can be regarded 
as a constant around the $\pi \Sigma N$ threshold. For instance, 
we have
\begin{eqnarray}   
S_{YN}^{{\rm con}}(E) &\approx& 
{\rm const.} \times \frac{1}{D^2(E)}
\label{eq:1/D^2(E)}
\end{eqnarray}
for the $Y+N$ decay spectrum. 

\Figuretable{FIG. 7}

\begin{figure}
  \begin{center}
\includegraphics[angle=90,width=0.80\linewidth]{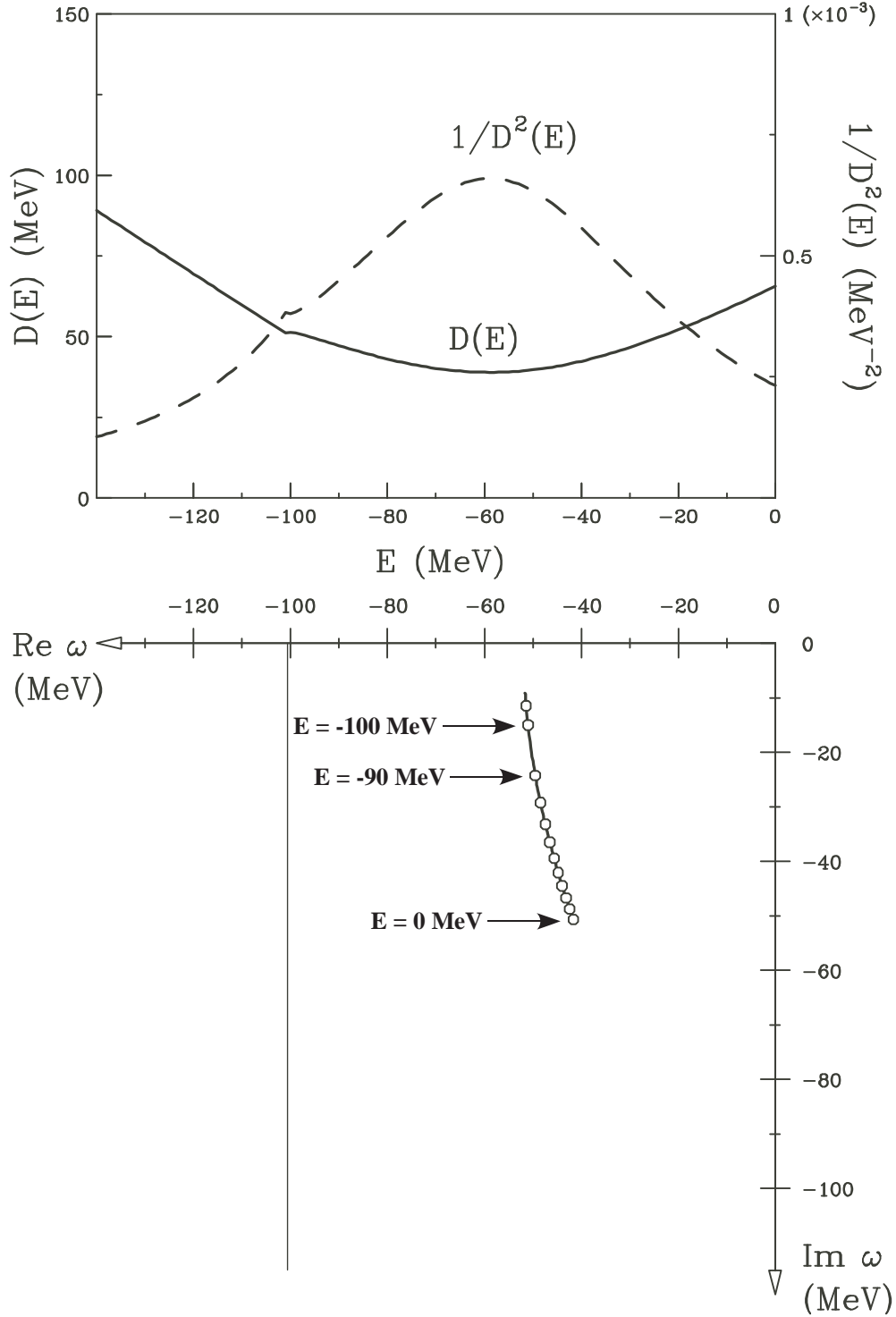}
\caption{\label{fig:7}
Pole trajectory of the $K^-pp$ state (bottom) and $D(E)$ (top) 
in the complex energy plane, 
in the case of potential B
($V_0 =$ $-$292 MeV and $W_0 =$ $-$107 MeV). 
The circles denote the pole positions at $\omega(E)$, 
which are drawn from $E = -110$ MeV to 0 MeV
in steps of 10 MeV.
The dashed curve denotes $1/D^2(E)$ which roughly represents 
the contribution of the pole in the spectrum.
  }
  \end{center}
\end{figure}

Now we consider the peak structure in the 
$[K^-pp] \to Y+ N$ decay spectrum  
obtained with $V_0=-292$ MeV and $W_0=-107$ MeV
for potential B.
In Fig.~\ref{fig:7}, we illustrate its pole trajectory of $\omega(E)$
as a function of $E$, together with $D(E)$ and $1/D^2(E)$. 
When $E$ is changed from 0 MeV to $-100$ MeV, 
the pole moves slowly from a point ($-42$ MeV, $-51$ MeV)
to a point ($-51$ MeV, $-15$ MeV), so that 
$D(E)$ works in an almost smooth function 
with the minimum value at $E \simeq -60$ MeV.
In this case, therefore, a clear peak in $1/D^2(E)$ 
is observed around $E \simeq -60$ MeV.
This peak is in good agreement with that of the
$Y + N$ decay spectrum in Fig.~\ref{fig:5}(b).
The shape of the spectrum is deviated from the standard BW form
due to the nature of the energy dependence of $D(E)$,  
whereas the position of this peak 
does not coincide with a point at $E=-B.E.=-45$ MeV.

\Figuretable{FIG. 8}
\Figuretable{FIG. 9}

\begin{figure}
  \begin{center}
\includegraphics[angle=90,width=0.80\linewidth]{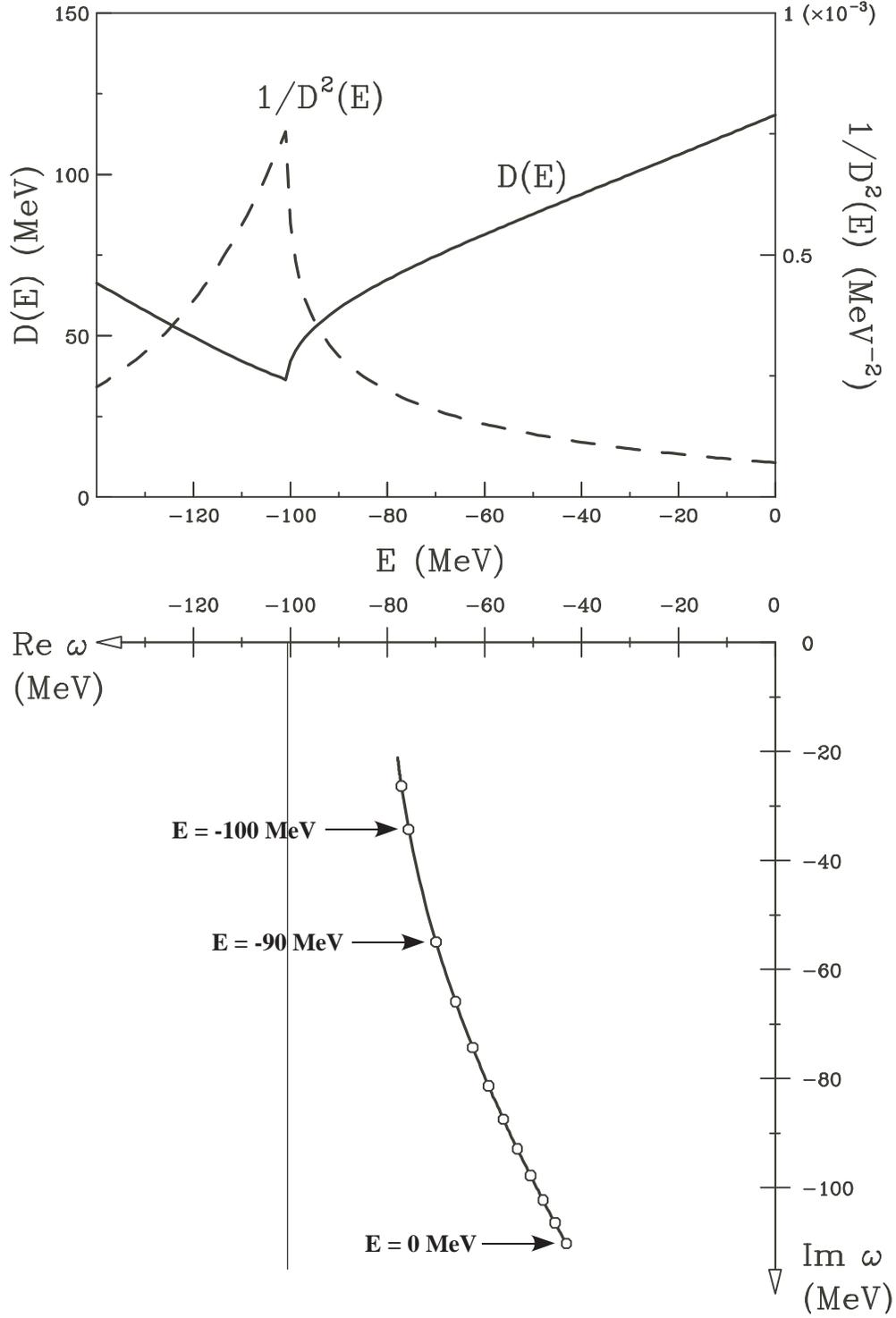}
\caption{\label{fig:8}
Pole trajectory of the $K^-pp$ state and $D(E)$ 
in the complex energy plane, 
in the case of potential C
($V_0 =$ $-$344 MeV and $W_0 =$ $-$203 MeV).
See also the caption of Fig.~\ref{fig:7}.
  }
  \end{center}
\end{figure}

\begin{figure}
  \begin{center}
\includegraphics[angle=90,width=0.80\linewidth]{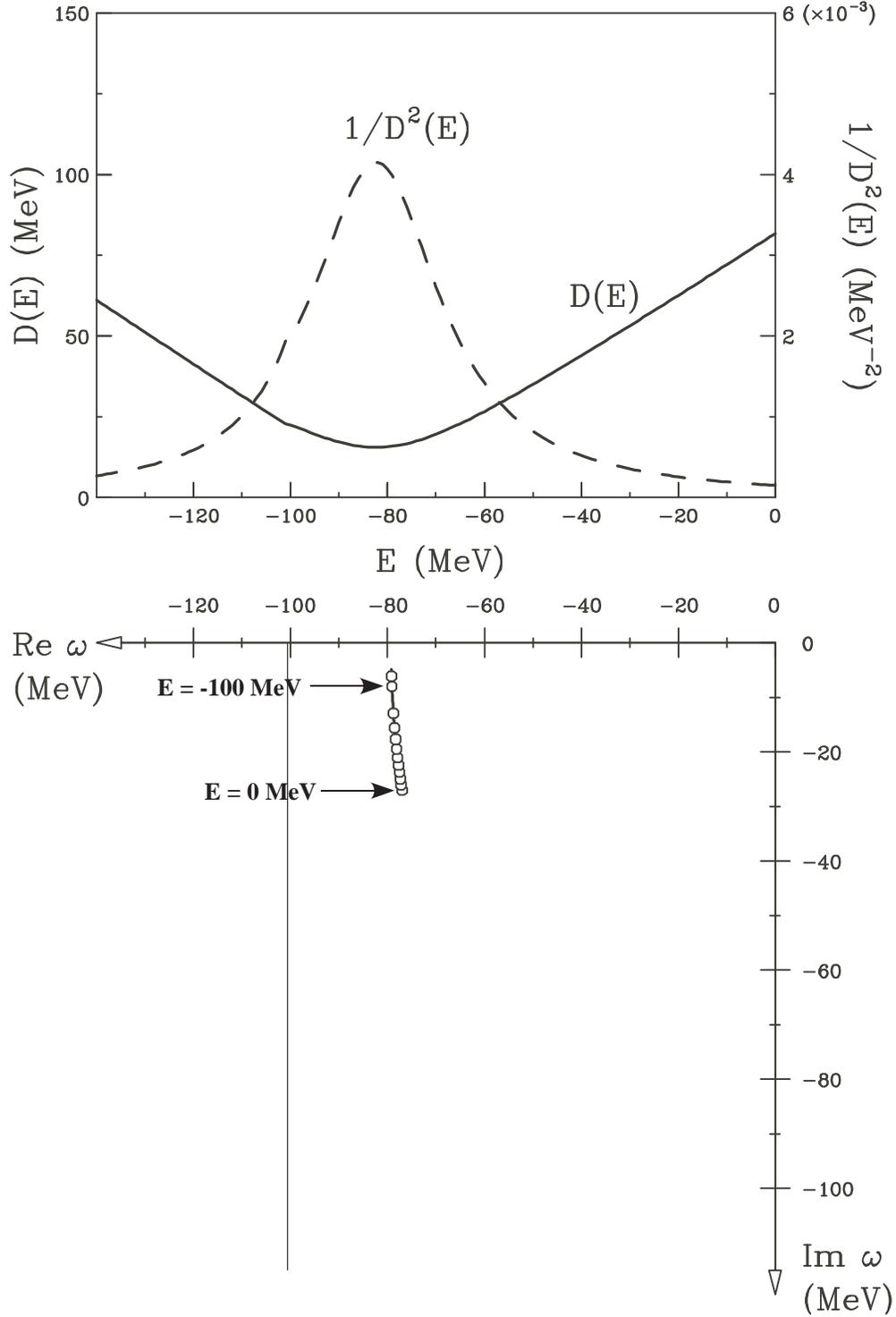}
\caption{\label{fig:9}
Pole trajectory of the $K^-pp$ state and $D(E)$ 
in the complex energy plane, 
in the case of potential C modified 
with an artificial narrow width 
($V_0 =$ $-$344 MeV and $W_0 =$ $-$47 MeV).
See also the caption of Fig.~\ref{fig:7}.
  }
  \end{center}
\end{figure}

\begin{figure}
  \begin{center}
\includegraphics[angle=90,width=0.80\linewidth]{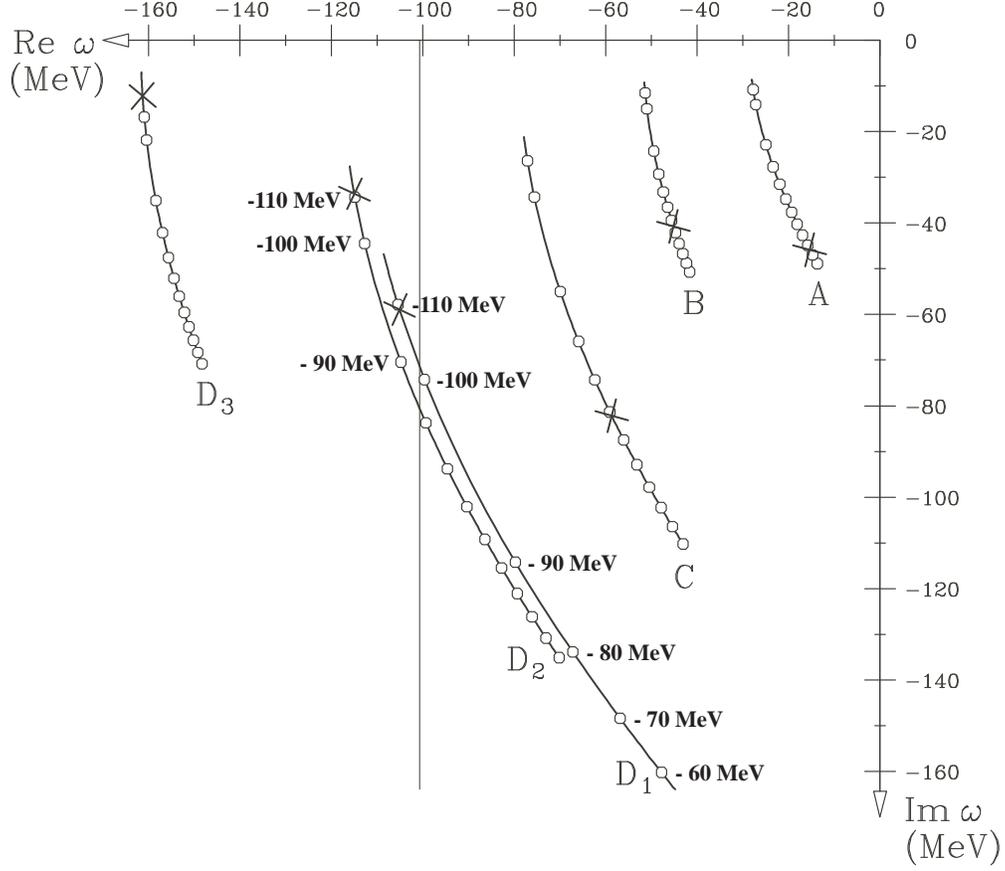}
\caption{\label{fig:10}
Pole trajectories of the $K^-pp$ state for potentials 
A, B, C, D$_1$, D$_2$ and D$_3$  
in the complex energy plane. 
The circles denote the pole positions at $\omega(E)$, 
which are drawn from $E = -110$ MeV to 0 MeV
in steps of 10 MeV.
The crosses denote positions at 
($-B.E.$, $-\varGamma/2$) of which values are given 
in Table \ref{tab:1} with these potentials.  
}
  \end{center}
\end{figure}

In Fig.~\ref{fig:5}(c), on the other hand, we have shown the cusp-like 
peak at the $\pi \Sigma N$ threshold 
in the $[K^-pp] \to Y + N$ decay spectrum 
for potential C. 
For understanding the appearance of such a cusp-like structure, 
we obtain a moving pole at $\omega(E)$ 
with $V_0 = -344$ MeV and $W_0 = -203$ MeV for potential C, 
as a function of $E$.
In Fig.~\ref{fig:8}, we illustrate the pole trajectory 
of their $\omega(E)$, $D(E)$ and $1/D^2(E)$. 
We find that the pole of $\omega(E)$ moves widely in the complex 
energy plane; when $E$ is changed from 0 MeV to $-100$ MeV, 
the pole of $\omega(E)$ moves from a point ($-43$ MeV, $-110$ MeV) 
to a point ($-76$ MeV, $-34$ MeV). 
For $E < E_{\rm th}(\pi \Sigma N)$, its pole remains around 
the point ($-77$ MeV, $-25$ MeV). 
It should be noticed that the minimum of $D(E)$ 
is realized at $E= E_{\rm th}(\pi\Sigma N)$ where 
$dD(E)/dE$ is singular.
In this case, therefore, a cusp-like peak in $1/D^2(E)$ is 
observed at the $\pi \Sigma N$ threshold. 
This shape agrees with that of the spectrum shown 
in Fig.~\ref{fig:5}(c). 
In order to see the effects of $W_0$, we also obtain a
trajectory of the moving pole with 
$V_0 = -344$ MeV and $W_0 = -47$ MeV, 
which corresponds to a specific case with an artificial narrow width. 
We confirm that the shape of its spectrum 
is identified as the BW form, as shown in Fig.~\ref{fig:9}.

Consequently, we recognize that the cusp-like structure 
can be described as behavior of the pole trajectory which is 
governed by the energy dependence of $U^{\rm opt}(E)$, as well as 
a clear peak with the BW form. 
The path of the trajectory for the moving pole 
in the complex energy plane
is determined by the values of $V_0$,  
and its moving range on the 
trajectory depends on the values of $W_0$.

\subsubsection{Pole trajectories by the $K^-$-``$pp$'' optical potentials}

\Figuretable{FIG. 10}

In Fig.~\ref{fig:10}, we show the pole trajectories 
of the $K^-pp$ bound state 
for potentials A, B, C, D$_1$, D$_2$ and D$_3$
in the complex energy plane. 
The strength parameters of ($V_0$,~$W_0$) characterize the shape 
structure of the $K^-pp$ state in
the spectrum with the $[K^-pp] \to Y + N$ decay from the two-nucleon 
$K^-$ absorption. 
For potential C, 
a cusp at the $\pi \Sigma N$ threshold appears clearly 
in the spectrum, as seen in Fig.~\ref{fig:5}(c), 
because the value of $D(E)$ at $E=E_{\rm th}(\pi \Sigma N)$ is 
much smaller than that of $\varGamma/2$ which is equivalent to 
the distance from the pole at a point ($-B.E.$,~$-\varGamma/2$) to 
the real axis.
For potential D$_2$, 
a steep step is observed at the $\pi \Sigma N$ 
threshold, as seen in Fig.~\ref{fig:5}(d); 
its yield is sharply cut down because its pole is 
rapidly moving above the $\pi \Sigma N$ threshold.

\Figuretable{FIG. 11}

One should be noticed that our $K^-$-``$pp$'' optical potentials, 
$U^{\rm opt}(E)$, are not derived from microscopic calculations, 
but are introduced phenomenologically.
To examine whether the potential has
the appropriate energy dependence or not, 
we evaluate the pole trajectory of a point $(-B.E., -{\varGamma}/2)$ 
in the complex energy plane when we change the value of $V_0$ 
in $U^{\rm opt}(E)$.

\begin{figure}[htb]
  \begin{center}
  \includegraphics[angle=90,width=0.80\linewidth]{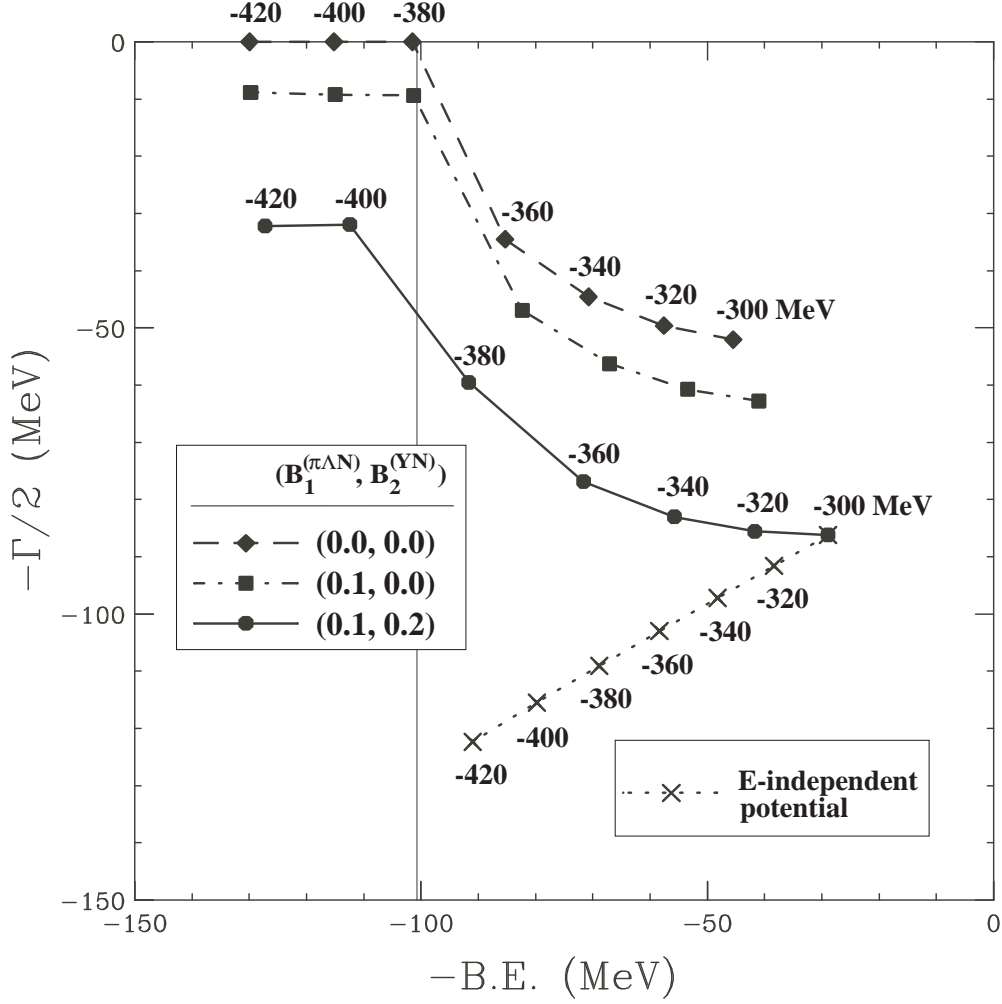}
  \caption{\label{fig:11}
The behavior of the moving pole at $(-B.E., -{\varGamma}/2)$ 
when the value of $V_0$ for the energy-dependent potential 
$U^{\rm opt}(E)$ is changed.
The diamonds, squares and circles denote 
the cases of 
($B_1^{(\pi \Lambda N)}$,~$B_2^{(Y N)}$) 
= (0.0,~0.0), (0.1,~0.0) and (0.1,~0.2) 
in $U^{\rm opt}(E)$ with $B_1^{(\pi \Sigma N)}=0.7$
and $W_0$ = $-$203 MeV, respectively.
The crosses denote the case of the energy-independent
potential $U^{\rm opt}_0$ with $W_0$ = $-$179 MeV.
The numbers attached to the symbols give the 
corresponding values of $V_0$.
}
  \end{center}
\end{figure}

In Fig.~\ref{fig:11}, we show the energy dependence of 
the pole trajectories on decay channels with $V_0=$
$(-300)\text{--}(-420)$ MeV, when we switch on/off each
$B^{(\pi \Lambda N)}_1$ and $B^{(Y N)}_2$
with $B^{(\pi \Sigma N)}_1 = 0.7$ and $W_0 = -203$ MeV
which corresponds to the imaginary part of the potential C.
If we consider only $B^{(\pi \Sigma N)}_1$,  
its width becomes smaller when the $-B.E.$
is close to $E_{\rm th}(\pi \Sigma N)$, and it finally becomes 0 
when $-B.E.$ is located below $E_{\rm th}(\pi \Sigma N)$.
This behavior seems to be qualitatively consistent with the result 
obtained from a $\bar{K}NN$-$\pi \Sigma N$ coupled-channel Faddeev 
calculation by Ikeda and Sato~\cite{Ike09}.
Even if $B^{(\pi \Lambda N)}_1$ and/or $B^{(Y N)}_2$ are switched on, 
the pole trajectory of the point ($-B.E.$,~$-{\varGamma}/2$) 
is not so changed quantitatively, except for an additional width.
But the pole trajectory for the energy-independent potential 
$U^{\rm opt}_0$ differs from that for $U^{\rm opt}(E)$;
$-{\varGamma}/2$ is almost proportional to $-B.E.$
Therefore, we believe that our $K^-$-``$pp$'' optical potential 
$U^{\rm opt}(E)$ has the desirable 
energy dependence which is expected from the coupled-channel 
Faddeev calculation, and that it is enough for us to discuss 
the shape of the spectrum with the $[K^-pp] \to Y+N$ decay process. 
For more quantitative argument, one should make 
the $\bar{K}NN$ single-channel effective potential,
in which the $\pi\Sigma N$ channel is eliminated in the 
$\bar{K}NN\text{--}\pi \Sigma N$ coupled channel scheme~\cite{Feshbach},
and compare it with our optical potential.
The investigation along this line will be discussed 
in future works.

\section{Discussions}
\label{Discuss}

\subsection{Cusp-like structure in the spectrum 
near the $\pi\Sigma N$ threshold}
\label{Condion_Cusp}

Recently, Akaishi $et$ $al.$~\cite{AKY08} have discussed
a cusp-like structure in the spectrum of 
the (in-flight $K^-$, $n$) reaction on a deuteron target, 
using a coupled-channel model with a separable potential. 
They have shown that the cusp-like structure at the $\pi \Sigma$
threshold can be also observed in the 
$[K^-p] \to \pi + \Sigma$ spectrum from the one-nucleon 
$K^-$ absorption. 
In Sect.~\ref{Results}, we have found a cusp-like 
structure at the $\pi \Sigma N$ threshold in 
the spectra with $[K^-pp] \to \pi + \Lambda + N$ 
and  $[K^-pp] \to Y + N$ decay precesses
if we use potential C.
The shape and magnitude of these spectra 
strongly depend on the pole trajectory of 
the $K^-pp$ bound state,
and are characterized  by 
the strength parameters of ($V_0$,~$W_0$).
It is worth examining the condition for ($V_0$,~$W_0$)
which gives the cusp-like structure at the $\pi \Sigma N$ threshold
within our optical potential, $U^{\rm opt}(E)$.

In this subsection, we focus on the spectra with 
$[K^-pp] \to \pi + \Sigma + N$ 
and $[K^-pp] \to Y + N$ decays, 
by artificially changing 
($V_0$,~$W_0$) in the following two cases: 
(i) ($B_1^{(\pi \Sigma N)}$,~$B_2^{(Y N)}$) = (0.7,~0.0)
which means only the $[K^-pp] \to \pi + \Sigma +N$ decay 
process; 
(ii) ($B_1^{(\pi \Sigma N)}$,~$B_2^{(Y N)}$) = (0.7,~0.2).
Here we omit $B^{(\pi \Lambda N)}_1$ for 
the $[K^-pp] \to \pi + \Lambda + N$ decay process
for simplicity, because $B^{(\pi \Lambda N)}_1$ 
operates similarly to $B^{(Y N)}_2$ in the spectrum, 
as discussed in Sect.~\ref{semi-exclusive}. 

\Figuretable{FIG.12}

\begin{figure}
  \begin{center}
\includegraphics[angle=90,width=0.7\linewidth]{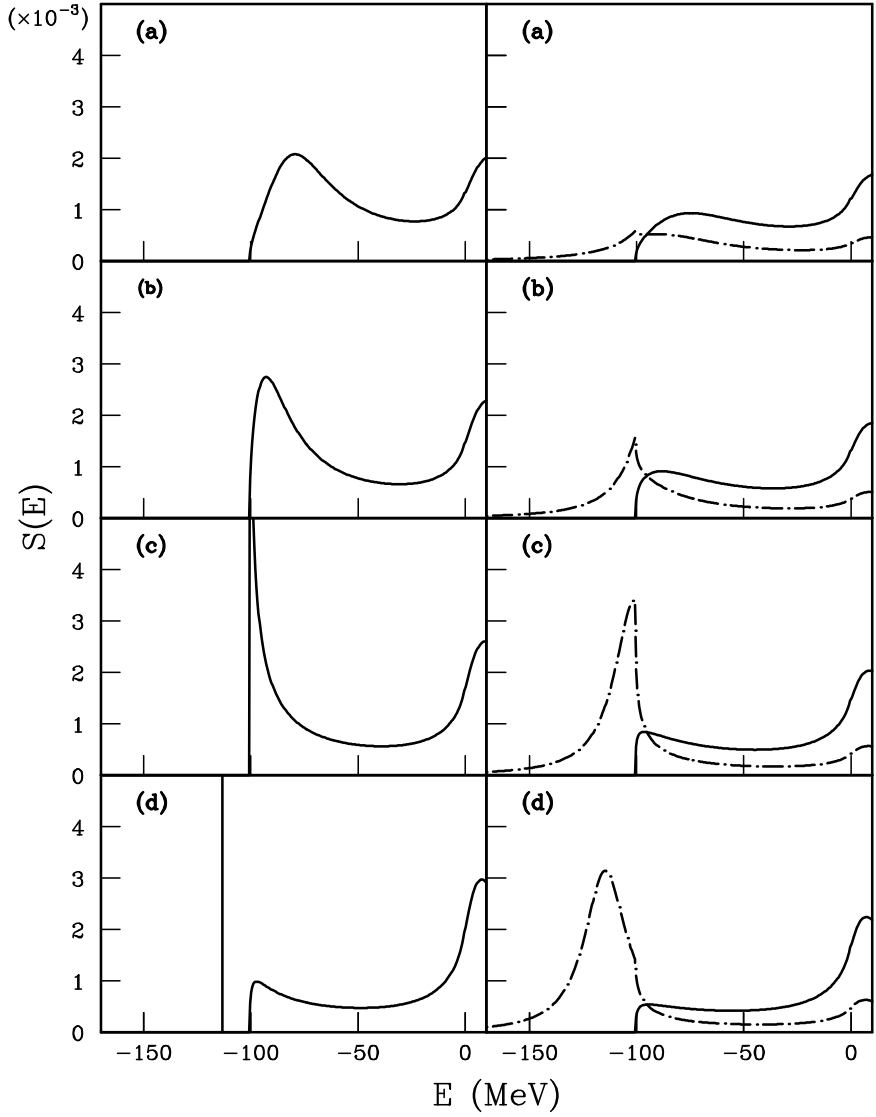}
\caption{\label{fig:12}
Behavior of the strength function $S(E)$ at $W_0=$ $-$107 MeV
when changing the value of $V_0$; 
(a) $V_0 =$ $-$340 MeV, (b) $-$360 MeV, (c) $-$380 MeV and (d) $-$400 MeV
in (left) the case of ($B^{(\pi \Sigma N)}_1$,~$B^{(Y N)}_2$) = (0.7,~0.0) 
and (right) the case of ($B^{(\pi \Sigma N)}_1$,~$B^{(Y N)}_2$) = (0.7,~0.2).
The solid and dashed curves denote $S_{\pi\Sigma N}^{{\rm con}}(E)$ 
and $S_{YN}^{{\rm con}}(E)$, respectively.
  }
  \end{center}      
\end{figure}

In Fig.~\ref{fig:12}, we show the behaviors of $S(E)$ 
in the spectra with $[K^-pp] \to \pi + \Sigma + N$ and 
$[K^-pp] \to Y + N$ decays at $-W_0$ = $107$ MeV, 
which corresponds to the imaginary part of potential B,  
by changing $-V_0=$ $340\text{--}400$ MeV.
For the case of (i), as seen in Fig.~\ref{fig:12}(left), 
the magnitude of the peak in $S_{\pi\Sigma N}^{{\rm con}}(E)$ 
grows at the $\pi \Sigma N$ threshold as increasing $-V_0$. 
When $-V_0 \simeq$ 380 MeV, the magnitude is at its maximum 
around $-B.E. \simeq E_{\rm th}(\pi \Sigma N)$.
When $-V_0 > 380$ MeV, the $K^-pp$ state must be bound below 
the $\pi\Sigma N$ threshold and its peak is located at $E$= $-B.E.$ 
Such a cusp-like peak in $S_{\pi\Sigma N}^{{\rm con}}(E)$ 
is quite similar to that obtained by Akaishi $et$ $al.$~\cite{AKY08}.

For the case of (ii), we show the behaviors of 
$S_{YN}^{{\rm con}}(E)$ and $S_{\pi\Sigma N}^{{\rm con}}(E)$
with $B_2^{(Y N)}=$ 0.2 in Fig.~\ref{fig:12}(right).
We find that there is the cusp-like structure in 
$S_{YN}^{{\rm con}}(E)$ at $-V_0\simeq$ $340\text{--}380$ MeV, 
and the asymmetric peak which is cut off sharply above the 
$\pi \Sigma N$ threshold appears in $S_{YN}^{{\rm con}}(E)$
when $-V_0 >$ 380 MeV, whereas 
there is no (cusp-like) peak in $S_{\pi\Sigma N}^{{\rm con}}(E)$.
Thus, we recognize that the cusp-like structure is 
observed in the $[K^-pp] \to Y + N$ decay spectrum rather than the
$[K^-pp] \to \pi + \Sigma + N$ decay one because of the existence 
of the $Y + N$ decay channel.

\Figuretable{FIG. 13}

\begin{figure}
  \begin{center}
\includegraphics[angle=90,width=0.7\linewidth]{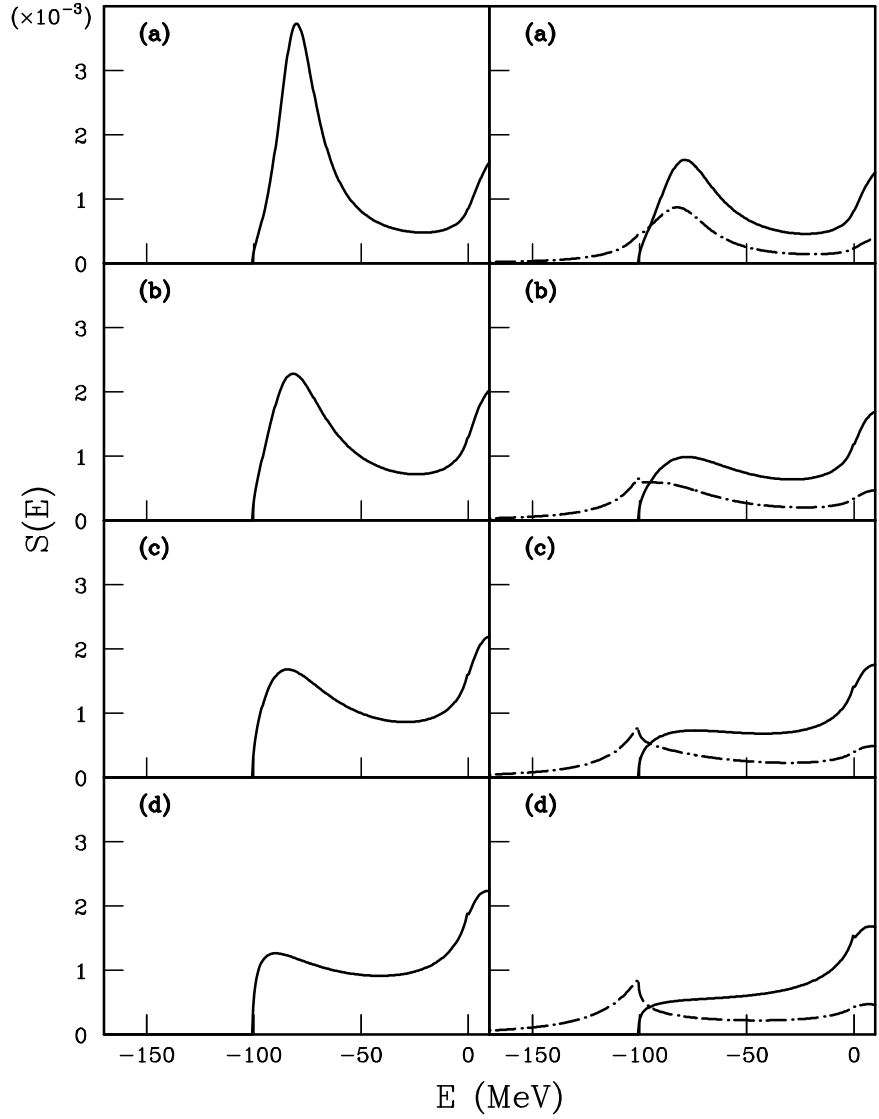}
\caption{\label{fig:13}
Behavior of the strength function $S(E)$ at $V_0=$ $-$344 MeV
when changing the value of $W_0$; 
(a) $W_0 =$ $-$60 MeV, 
(b) $-$100 MeV, (c) $-$140 MeV and (d) $-$200 MeV.
See also the caption in Fig.~\ref{fig:12}.
  }
  \end{center}      
\end{figure}

In Fig.~\ref{fig:13}, we examine the behaviors of $S(E)$ 
in the spectra with $[K^-pp] \to \pi + \Sigma + N$ and 
$[K^-pp] \to Y + N$ decays at $-V_0 =$ 344 MeV,
which corresponds to the real part of potential C,
by changing $-W_0=$ $60\text{--}200$ MeV.
In the case of (i), we obtain that 
a clear peak near $E \simeq$ $-80$ MeV in 
$S_{\pi\Sigma N}^{{\rm con}}(E)$  becomes broad
as $-W_0$ increases, 
as shown in Fig.~\ref{fig:13}(left). 
In the case of (ii), 
we also find that a clear peak is located 
near $E \simeq$ $-80$ MeV in both of 
$S_{\pi\Sigma N}^{{\rm con}}(E)$ and 
$S_{YN}^{{\rm con}}(E)$ at $-W_0=$ $60$ MeV,
as shown in Fig.~\ref{fig:13}(right).
As increasing $-W_0$, 
$S_{\pi\Sigma N}^{{\rm con}}(E)$ is gradually reduced,
but $S_{YN}^{{\rm con}}(E)$ is gradually enhanced just 
at the $\pi \Sigma N$ threshold.
Thus it grows up a threshold-cusp in the spectrum. 

Therefore, we have the cusp-like structure 
in the spectrum under the conditions 
that $-B.E.$ is close to and above the $\pi \Sigma N$ 
threshold energy and $\varGamma$ is considerably large.
The corresponding strength parameters are roughly 
estimated as $-V_0 =$ $330\text{--}380$ MeV 
and $-W_0 \geq$ $100\text{--}120$ MeV,
which are found in such as potential C given 
in Table~\ref{tab:1}.
The cusp-like structure is the unique signal for 
evidence of the deeply bound $K^- pp$ state in 
the $[K^-pp] \to Y + N$ decay spectrum.

\subsection{Dependence of the spectrum on the branching 
rate of $B^{(YN)}_2$} 
\label{2NAdependence}

The shape of the semi-exclusive 
$K^-$ conversion spectrum including  
the $[K^-pp] \to Y + N$ decay process
is very important 
to extract the structure of the $K^-pp$ state, 
e.g., the potential strengths of ($V_0$,~$W_0$), 
rather than that of the spectrum including the 
$[K^-pp] \to \pi + \Sigma + N$ decay process.
In our calculations, we assumed the branching rate of 
$B_2^{(Y N)}=0.2$ in the $K^-pp$ decay processes.
This value is often used in previous 
works for heavier targets~\cite{Kis05,Yam06,Mar05,Gaz07} 
but it is experimentally unknown 
in the $K^-$ absorption on $^3$He in flight.
In terms of the $K^-$ absorption on $^4$He at rest, 
the early data of the helium bubble chamber experiment~\cite{Kat70} 
suggested that the ratio of the two-nucleon $K^-$ absorption 
to all the $K^-$ absorption processes amounts to 16 \%, whereas
its value depends on atomic orbits where $K^-$ is absorbed through 
atomic cascade processes~\cite{Ona89,Yam06}. 
A recent analysis of the $K^-$ absorption on $^4$He at rest~\cite{Suz08} 
also calls for reexamination of $B_2^{(Y N)}$ experimentally. 
To determine the value of  $B_2^{(Y N)}$ in the 
$K^-$ absorption in flight, 
we need more investigations on $B_2^{(Y N)}$ 
experimentally and theoretically.

\Figuretable{FIG. 14}

\begin{figure}
  \begin{center}
\includegraphics[angle=90,width=0.5\linewidth]{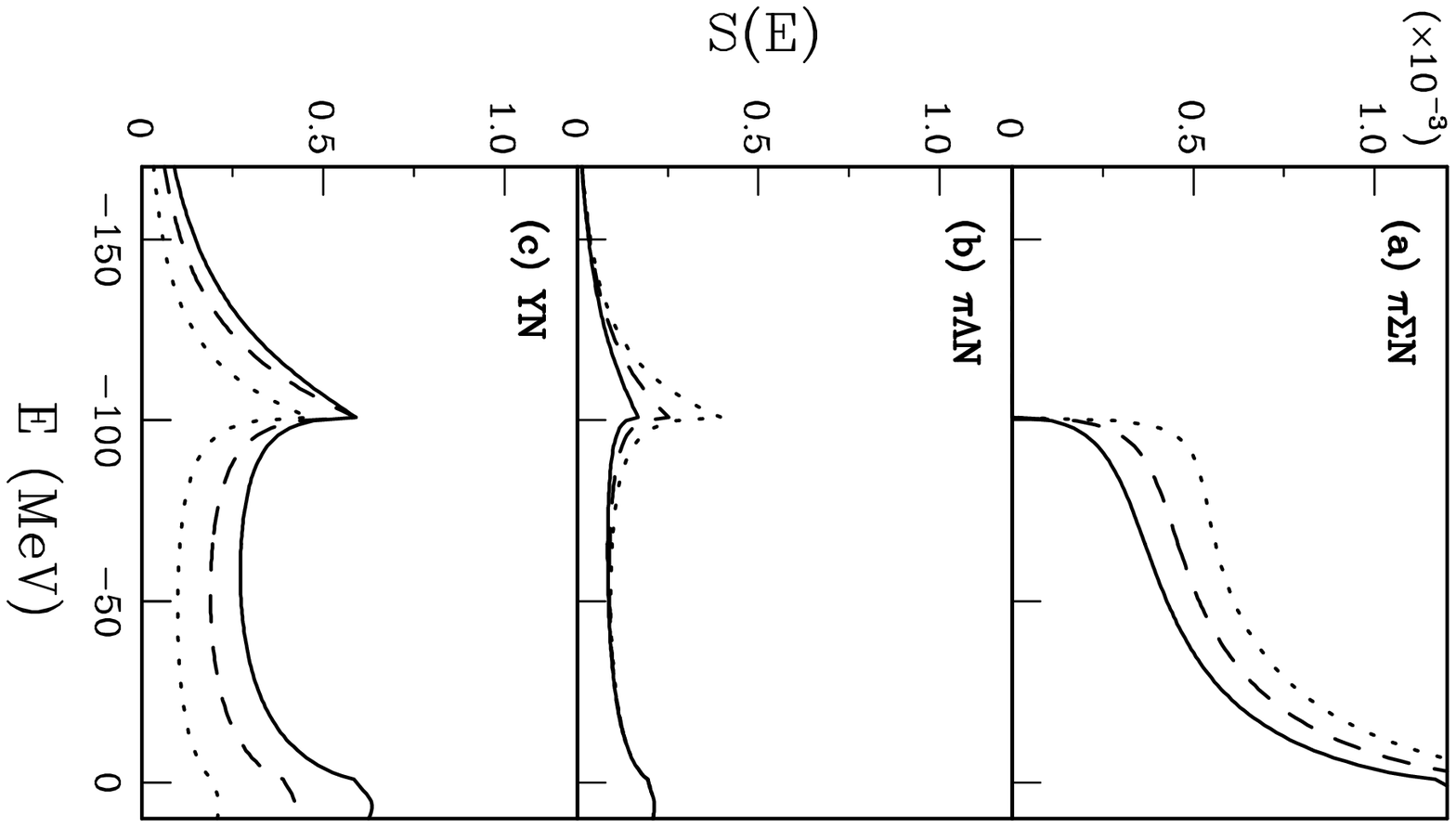}
\caption{\label{fig:14}
Behavior of the calculated strength functions $S(E)$
for (a) $[K^-pp] \to \pi + \Sigma + N$,
(b) $[K^-pp] \to \pi + \Lambda + N$
and (c) $[K^-pp] \to Y + N$ decay processes. 
Here potential C
with $B_1^{(\pi \Lambda N)}=0.1$ is used.
The dotted, dashed and solid curves denote the spectra 
for ($B_1^{(\pi \Sigma N)}$,~$B_2^{(Y N)}$)=
(0.8,~0.1), (0.7,~0.2) and (0.6,~0.3), respectively.
  }
  \end{center}
\end{figure}

As the first step forward the investigations, 
we attempt to calculate the strength function $S(E)$ in potential C,
in order to check the sensitivity of semi-exclusive spectra 
on the value of $B_2^{(Y N)}$. 
In Fig.~\ref{fig:14}, we demonstrate the dependence of the spectra 
on the values of $B_2^{(Y N)}$ when changing 
$B_2^{(Y N)}=$ $0.1\text{--}0.3$.
For the $[K^-pp] \to \pi + \Sigma + N$ and  
$[K^-pp] \to \pi + \Lambda + N$ decay spectra, 
each magnitude is reduced as $B_2^{(Y N)}$ increases, 
as shown in Figs.~\ref{fig:14}(a) and (b).
On the other hand, the $[K^-pp] \to Y + N$ decay spectrum
is enhanced as $B_2^{(Y N)}$ increases, 
as shown in Fig.~\ref{fig:14}(c).
The shape of these spectra is scarcely modified 
by a small change of $B_2^{(Y N)}$.
Thus, the detailed values of the branching rates
have an influence only on the relative magnitude 
of each decay spectrum, without changing the nature 
of the $K^-pp$ formation signal.
We stress that it is important to compare the shapes of 
the calculated spectra with those of the measured ones.
This detailed comparison provides valuable information 
on $B_2^{(\pi \Sigma N)}$, $B_2^{(\pi \Lambda N)}$ and $B_2^{(Y N)}$, 
as well as on the binding energy and width of the $K^- pp$ state.

\subsection{The spectrum near the $K^- +p + p$ threshold} 

In Fig.~\ref{fig:5}(a), we have shown partial contributions 
of the spectra with potential A. 
The spectra have no clear peak because 
the pole for potential A has a large width 
and is located near the $K^- +p+p$ threshold.
In order to extract information on the 
$K^-$-``$pp$'' potential from their spectral shape, 
one needs to consider the effect of the $K^- +p+p$ threshold 
beyond the ``$pp$'' core assumption.
If $``pp\mbox{''} \to p + p$ degree of freedom 
is taken into account, a QF $\Lambda(1405)$ formation via 
$[K^-pp] \to ``K^-p\mbox{''} + p \to \Lambda(1405) + p$
would be important rather than 
$[K^-pp] \to K^- + p + p$ break-up processes. 
The spectrum of such a QF $\Lambda(1405)$ formation stands up 
from $E \simeq$ ($-$10)-($-$20) MeV below the $K^-+p+p$ threshold, 
which depends on the $\Lambda(1405)$ mass as a $\bar{K}\text{--}N$ 
quasibound state.
In the case of $B.E.\simeq$ 20 MeV obtained in potential A,  
therefore, it may be necessary to estimate the contribution of the QF 
$\Lambda(1405)$ spectrum which contaminates 
into the $K^-pp$ formation spectrum near the $K^-+p+p$ threshold.

\section{Summary and Conclusion}
\label{Summary}

We have examined 
the inclusive and semi-exclusive spectra in the 
$^3$He(in-flight $K^-$, $n$) reaction at $p_{K^-}$ = 1.0 GeV/c 
and $\theta_{\rm lab}=$ 0$^\circ$ for the forthcoming J-PARC E15 experiment.
We have discussed these spectra with the energy-dependent $K^-$-``$pp$'' optical
potentials $U^{\rm opt}(E)$, based on the results of the binding 
energies and widths of the $K^- pp$ (unstable) bound states 
in several predictions or candidates. 
To understand the peak structure in the spectrum,
we have investigated the trajectory of the moving pole of 
the $K^-pp$ bound state in the complex energy plane, 
and the behavior of the corresponding strength function 
by changing the strength parameters ($V_0$,~$W_0$) of 
$U^{\rm opt}(E)$ systematically.
The calculated spectrum predominately has the bound state with 
$I=$ 1/2, $J^\pi=$ 0$^-$ and $L = 0$ 
in the $K^-pp$ bound region below the $K^- +p+p$ 
threshold, whereas the continuum states with $L \geq 1$ occur
in the QF region.
We have shown that the (in-flight $K^-$,~$n$) reaction 
on the $s$-shell nuclear targets such as $^3$He 
provides an advantage to produce the deeply-bound 
$K^-$ nuclear state with $L=0$. 
The results are summarized as follows:
\begin{itemize}
\item[(i)]
The clear peak appears below the $\pi\Sigma N$ threshold
in the spectrum with the $[K^-pp] \to Y + N$ decay from the 
two-nucleon $K^-$ absorption 
as evidence of the $K^- pp$ bound state, 
within $-V_0 >$ 380 MeV, 
as the case of potentials D.

\item[(ii)]
The cusp-like structure appears at the $\pi\Sigma N$ 
threshold in the $[K^-pp] \to Y + N$ decay spectrum
within $-V_0 \simeq$ $330\text{--}380$ MeV and $-W_0 >$ $\sim$110 MeV, 
rather than in the $[K^-pp] \to \pi + \Sigma + N$ decay 
spectrum, 
as the case of potential C.

\item[(iii)]
The distinct peak in the $[K^-pp] \to Y + N$ and 
$[K^-pp] \to \pi + \Sigma + N$ decay spectra 
is observed as clear evidence of the $K^- pp$ bound state 
within $-V_0 \simeq$ $200\text{--}330$ MeV and 
$-W_0 <$ $\sim$110 MeV such as 
potential B, 
whereas no clear peak is observed in these spectra 
even if the $K^- pp$ bound state exists 
within $-V_0 \simeq$ $200\text{--}330$ MeV and $-W_0 >$ $\sim$110 MeV
such as potential A.
\end{itemize}

In conclusion, 
the $^3$He(in-flight $K^-$, $n$) spectrum including the $[K^-pp] \to Y +N$ 
decay process from the two-nucleon $K^-$ absorption
provides evidence of the $K^- pp$ bound state 
to identify itself as the appropriate $K^-$-``$pp$'' potential
with the help of the trajectory of its moving pole in the complex energy 
plane.
If any of the experimental observations of 
DISTO~\cite{DISTO08}, FINUDA~\cite{FINUDA05} and OBELIX~\cite{OBELIX07} 
indicates evidence of the $K^- pp$ bound state,
its corresponding peak should appears below
the $\pi\Sigma N$ threshold in the J-PARC E15 spectrum.
Otherwise, we will realize that these experimental data are all incorrect.
Moreover, the cusp-like structure is the unique signal of 
the $K^- pp$ formation, as well as the peak structure.
This phenomenology suggests the possibility of
observing the cusp-like structure obtained 
by the deep potential with strong absorption 
($-V_0 =$ $330\text{--}380$ MeV, $-W_0 >110$ MeV), 
as predicted by Shevchenko $et$ $al.$~\cite{She07}.
If a cusp-like structure is observed, 
a precise comparison between theoretical and experimental 
spectra is required to extract the binding energy and width 
of the $K^- pp$ state, as well as the analysis of the spectrum 
in which the clear peak is observed.
To get more quantitative results on the cusp-like or peak 
structure, a full microscopic calculation between $\bar{K}NN$
and $\pi Y N$ channels would be required beyond 
our optical potential models.
Nevertheless, we believe that our calculations
lead to a good insight for qualitative understanding 
the spectrum of the deeply-bound $K^-pp$ state.

\begin{acknowledgments}

We acknowledge Prof. M. Iwasaki, Dr. H. Ohta, Dr. H. Ohnishi
and Dr. T. Suzuki for discussions about a plan of the J-PARC E15 
experiment. We would like to thank Prof. Y. Akaishi for 
many valuable discussions and suggestions.
This work is supported by Grants-in-Aid for Scientific Research on
Priority Areas (Nos. 17070002, 17070007 and 20028012). 

\end{acknowledgments}

\clearpage

\begin{table*}[bth]
\caption{
\label{tab:1}
Parameters of the real and imaginary strengths, $V_0$ and $W_0$, 
of the $K^-$-``$pp$'' optical potentials $U^{\rm opt}(E;{\bm r})$ 
for the $I=1/2$, $J^\pi=0^-$ bound state in Eq.(\ref{eq:E-dep_opt}). 
The range parameter is $b=$ 1.09 fm.
The branching rates of the one-nucleon $K^-$ absorption process
are taken to be $B_1^{(\pi \Sigma N)} =0.7$ and 
$B_1^{(\pi \Lambda N)} =0.1$, respectively, and the branching rate of 
the two-nucleon $K^-$ absorption process is $B_2^{(Y N)}=0.2$.
The values in the brackets denote for the imaginary parts 
of the energy-independent 
potentials $U^{\rm opt}_0({\bm r})$. The unit of all values is MeV. 
}
\begin{ruledtabular}
\begin{tabular}{lccccccl}
Potentials  &   $V_0$   & $W_0$
& \multicolumn{2}{l}{without $B_2^{(Y N)}$}
& \multicolumn{2}{l}{with $B_2^{(Y N)}$}&  Refs. \\
\cline{4-5} \cline{6-7}
&            & 
& $B.E.$\footnotemark[1] & ${\varGamma}$\footnotemark[2] 
& $B.E.$\footnotemark[1] & ${\varGamma}$\footnotemark[3] &  \\
\hline
\quad A  &  $-$237 & $-$128 ($-$120) & 21  &  70  & 15  &  92 & DHW~\cite{Dot08}\\
\quad B &  $-$292 & $-$107 ($-$86)  & 48  &  61  & 45  &  82 & YA~\cite{YA02} \\
\quad C  &  $-$344 & $-$203 ($-$147) & 70  & 110  & 59  & 164 & SGM~\cite{She07} \\
\quad D$_1$   &  $-$399 & $-$372 ($-$86)  & 114 &  34  & 105 & 118 & DISTO~\cite{DISTO08} \\
\quad D$_2$  &  $-$404 & $-$213 ($-$47)  & 118 &  19  & 115 &  67 & FINUDA~\cite{FINUDA05} \\
\quad D$_3$  &  $-$458 & $-$82 ($-$13)   & 162 &   5  & 161 &  24 & OBELIX~\cite{OBELIX07} \\
 \end{tabular}                                  
\end{ruledtabular}        
                         
\footnotetext[1]
{Binding energy of the $K^-pp$ bound state 
measured from the $K^- +p +p$ threshold.}
\footnotetext[2]
{Width from the one-nucleon $K^-$ absorption processes.}
\footnotetext[3]
{Total width from the one- and two-nucleon $K^-$ absorption processes.}
\end{table*}

%

%

%

%

\begin{thebibliography}{00}
\bibitem{NoteAdd}
In Note added in proof of Ref.~\cite{Yam08}, 
it is commented that the bound states with $L \geq 1$ should not be 
interpreted as the three-body $K^-pp$ state,
but could be $\Lambda$(1405) formed by the two-body $K^-p$ state.
\end{thebibliography}
\end{document}